\documentclass[journal]{IEEEtran} 
\PassOptionsToPackage{margin=0.75in}{geometry}
\usepackage[utf8]{inputenc}
\usepackage[english]{babel}
\usepackage{lettrine}
\usepackage[nottoc]{tocbibind}

\usepackage{hyperref}
\usepackage{lettrine}
\usepackage{amsmath}
\usepackage{booktabs}
\usepackage{mathtools}
\usepackage{amssymb,epsfig,multirow}
\usepackage{amsbsy}
\usepackage{amsthm}
\usepackage{titlesec}

\usepackage{graphicx}
\usepackage{authblk}
\usepackage{geometry}
\usepackage{url}
\usepackage{float}
\usepackage{color}
\usepackage{enumerate}
\usepackage{version}
\usepackage[linesnumbered,lined,boxed,commentsnumbered]{algorithm2e}
\usepackage{tikz}
\usetikzlibrary{arrows.meta}
\usepackage{tkz-euclide}
\usepackage{siunitx}
\usepackage{subcaption}
\usepackage{algpseudocode}
\usepackage{pifont}
\usetikzlibrary{shapes,snakes}
\newtheorem{lem}{Lemma}[section]
\newtheorem{defn}{Definition}[section]
\newtheorem{thm}{Theorem}[section]
\newtheorem{cor}{Corollary}[section]
\newtheorem{prop}{Proposition}[section]

\newtheorem{assum}{Assumption}[section]

\newtheorem{rem}{Remark}[section]
\newtheorem{obser}{Observation}[section]

\newcommand{\C}{\mathcal{C}}
\newcommand{\A}{\mathcal{A}}
\newcommand{\oo}{\mathcal{O}}
\newcommand{\F}{\mathcal{F}}

\title{Constrained Multi-Agent Path Finding on Directed Graphs}

\author{S. Ardizzoni, L.Consolini, M.Locatelli, I. Saccani
  \thanks{All authors are with the Dipartimento di Ingegneria
  	e Architettura, Universit\`a degli Studi di Parma,
  	Parco Area delle Scienze 181/A, 43124 Parma, Italy. E-mails:
  	{\tt\footnotesize \{stefano.ardizzoni, luca.consolini, marco.locatelli, irene.saccani
  		\}@unipr.it}}%
}

\date{}
\begin{document}
\maketitle
\begin{abstract}
We discuss $\C$-MP and $\C$-MAPF, generalizations of the classical
Motion Planning (MP) and Multi-Agent Path Finding (MAPF) problems on a directed graph $G$. Namely, we enforce an upper bound on the number of agents that occupy each member of a family of vertex subsets. For instance, this constraint allows maintaining a safety distance between agents. We prove that finding a feasible solution of $\C$-MP and $\C$-MAPF is NP-hard, and we propose a reduction method to convert them to standard MP and MAPF. This reduction method consists in finding a subset of nodes $W$ and a reduced graph $G_W$, such that 
a solution of MAPF on $G_W$ provides a solution of $\C$-MAPF on $G$. Moreover, we study the problem of finding $W$ of maximum cardinality, which is strongly NP-hard.
%have been widely studied in literature. However, classic MP and MAPF assume
%that an agent occupies only a single location at any
%given time, and that each node of the graph is equivalent to the others, without any particular constraint. This limits their applicability
%in many real-world domains. In this paper, we formalize and study $\C$-MP and $\C$-MAPF,
%i.e., MP and MAPF with additional constraints. Given a directed graph $G=(V,E)$ we represent the problem constraints as a set of pairs $(W,k)$, where $W \subset V$ and $k \in \mathbb{N}$. At any time, and for each pair $(W,k)$, we require that the nodes in $W$ contain at most $k$ agents. These additional constraints allow to take into account the area occupied by each agent, to maintain a safety distance between agents, and to fulfill some traffic rules within a warehouse. We prove that finding a feasible solution of $\C$-MP and $\C$-MAPF is NP-hard, and we propose a reduction method to convert them to standard MP and MAPF. This reduction method consists in finding a subset of nodes $W$ and a \textit{reduced} graph $G' = (W, E_W)$ such that 
%a solution of MAPF on $G'$ provides a solution of $\C$-MAPF on $G$. Moreover, we study the problem of finding $W$ of maximum cardinality, which is strongly NP-hard.
\end{abstract}

%{\bf Keywords:} Multi-Agent path finding, motion planning, complexity

\section{Introduction}
We discuss generalizations of two classical problems, related to the reconfiguration of set of agents, moving along the edges of a directed graph. The first, 
Multi-Agent Path Finding (MAPF in what follows), consists in finding a sequence of movements that re-positions all agents to assigned target nodes, avoiding collisions. The simpler Motion Planning (MP in what follows) consists in computing a sequence of movements for all agents, such that a single marked agent reaches a desired target node. That is, in MP, all non-marked agents are obstacles that need to be moved out of the way to re-position the marked one.

Our interest in MP and MAPF comes from the management of fleets of automated guided vehicles (AGVs). AGVs move items between different locations in a warehouse. Each AGV follows a predefined path, that connects locations in which items are stored or processed. AGVs are equivalent to agents in MAPF and MP. 
Note that, in industrial applications, multiple AGVs are able to move at the same time. On the other hand, as commonly made in MAPF and MP, we restrict the agents to move one at a time. This restriction is not so relevant since, in this paper, we focus on finding a feasible solution for MAPF and MP.
After a feasible solution is found, one could optimize it by local search procedures that allow simultaneous agents motions. However, this is not the focus of the present paper.

 %However, in congested situations AGVs are constrained to move one at a time in order to avoid deadlock situations and this leads to the solution of MP and MAPF problems. 

\begin{figure}[h!]
 \centering
 \begin{subfigure}[b]{0.4\textwidth}	
 	\begin{tikzpicture}%[scale=0.45]
 		
 		\node[shape=circle,draw=red,line width=1mm] (B) at (2,0) {1};
 		\node[shape=circle,draw=green,line width=1mm] (C) at (6,2) {3};
 		
 		\node[shape=circle,draw=green,line width=1mm] (D) at (3.5,2) {2};
 		
 		\node[shape=circle,draw=red,line width=1mm] (E) at (7.5,0) {4};
 		
 		%\node[shape=circle,draw=black] (G) at (3.5,-2) {6};
 		
 		\node[shape=circle,draw=black] (H) at (4.75,-2) {5};
 		
 		%\path [->] (C) edge[bend left=20]  (H);
 		
 		\path [->] (D) edge[bend left=20]  (C);

 		\path [->] (B)  edge[bend left=30]   (D);

 		\path [->] (C) edge[bend left=20]  (E);
 		
 		\path [->] (E)  edge[bend left=40]   (H);
 		%\path [->] (E)  edge[bend right=40]   (G);
 		
 		%\path [->] (G)  edge[bend left=30]   (B);
 		\path [->] (H)  edge[bend left=40]   (B);

 	\end{tikzpicture}
 	\caption{$D=(V,E), \quad \C=\{(\{1,4\},1), (\{2,3\},1)\}$.}
 	\label{ex0}
 \end{subfigure}
 \medskip
 
 \begin{subfigure}{0.4\textwidth}
 	
 	\begin{tikzpicture}%[scale=0.45]
 		
 		\node[shape=circle,draw=red,line width=1mm] (B) at (2,0) {1};
 		\node[shape=circle,draw=green,line width=1mm] (C) at (6,2) {3};
 		
 		\node[shape=circle,draw=green,line width=1mm] (D) at (3.5,2) {2};
 		
 		\node[shape=circle,draw=red,line width=1mm] (E) at (7.5,0) {4};
 		
 		%\node[shape=circle,draw=black] (G) at (3.5,-2) {6};
 		
 		\node[shape=circle,draw=black] (H) at (4.75,-2) {5}; 		
 		
 		\path [->] (C) edge[bend left=20]  (H);
 		
 		\path [->] (D) edge[bend left=20]  (C);
 		%\path [->] (A)  edge[bend right=60]  node[above] {$0$} (D);
 		%\path [->] (A)  edge[bend right=80]  node[above] {$0$} (E);
 		
 		\path [->] (B)  edge[bend left=30]   (D);
 		%\path [->] (B)  edge[bend left=80]  node[above] {$\beta_1$} (E);

 		\path [->] (C) edge[bend left=20]  (E);
 		
 		\path [->] (E)  edge[bend left=40]   (H);
 		%\path [->] (E)  edge[bend right=40]   (G);
 		
 		%\path [->] (G)  edge[bend left=30]   (B);
 		\path [->] (H)  edge[bend left=40]   (B);

 	\end{tikzpicture}

 	\caption{$D'=(V,E'), \quad \C=\{(\{1,4\},1), (\{2,3\},1)\}$.}
 	
 	\label{ex}
\end{subfigure}

\medskip

\caption{Examples of $C$-MAPF instances.}
\end{figure}
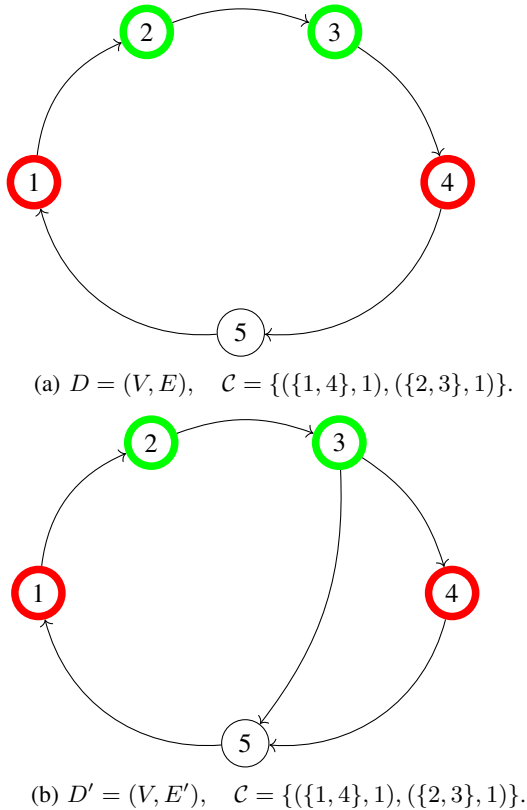

MP and MAPF have been widely studied in literature (\cite{MAPF,WuGru09,feasibility}). Researchers proposed various optimal or suboptimal algorithms (see, for instance, \cite{diSC,diBOX,cbs,hca,Tass,PMT}).
%Despite this, traveling on directed graphs is relevant in navigation domains, such as path finding in
%games, and asymmetric communication networks. \cite{diBOX} is the first work focused on directed graphs and presents $diBOX$, an algorithm that  computes in polynomial time suboptimal solutions  in the specific case of strongly biconnected digraphs. It is proved that this path solver is complete for instances of $MAPF$ with at least two holes. 
As said, we focus on a variant of MP and MAPF, where we take into account additional constraints.  
Given a directed graph $G=(V,E)$, where $V$ is the vertex set and $E$ the edge set, we represent the problem constraints as a set of pairs $(W,k)$, where $W \subset V$ and $k \in \mathbb{N}$. At any time, and for each pair $(W,k)$, we require that the nodes in $W$ contain at most $k$ agents.
For instance, consider the digraphs in Figures~\ref{ex0}-\ref{ex}, where the constraints are the two pairs $\left(\{2,3\},1\right)$ and 
 $\left(\{1,4\},1\right)$. In this way, at any time, we require that at most one agent is present in vertex subsets $\{2,3\}$ and $\{1,4\}$. In practice, this can be due to the fact that vertices $\{2,3\}$ and $\{1,4\}$ are too close to be occupied by two agents at the same time.
 In what follows, we denote by $\C$-MP and $\C$-MAPF the variants of MP and MAPF obtained by imposing constraints of this type.
 
With a suitable choice of pairs $(W,k)$, these additional constraints allow:
\begin{itemize}
\item taking into account the area occupied by each agent.
Even if it is common to assume that each agent occupies a single node, this is often not true. For instance, industrial AGVs are often quite large, and close nodes cannot be occupied at the same time
(see also the discussion of the previous example);
\item maintaining a safety distance between agents;
\item fulfilling some traffic rules within a warehouse, imposing a maximum number of agents in given areas. This can be due to constraints on floor load capacity, or to restrict the number of agents in areas shared with human workers.
\end{itemize} 
 % We point out that, for the sake of generality, later on constraints will be defined differently, although in practical contexts the above definition is sufficient.
%For example, consider the digraphs represented in Figures~\ref{ex0}, \ref{ex}, with the requirement that subsets $\{2,3\}$ and $\{1,4\}$ do not contain more th

\begin{figure}[h!]
\centering
\begin{subfigure}[b]{0.45\textwidth}
	%\centering
	\begin{tikzpicture}%[scale=0.4]

		\node[shape=circle,draw=green,line width=1mm] (D) at (3.5,2) {2};
		
		\node[shape=circle,draw=red,line width=1mm] (E) at (7.5,0) {4};

		\path [->] (E) edge[bend left=20]  (D);
		
		\path [->] (D) edge[bend left=20]  (E);
		%\path [->] (A)  edge[bend right=60]  node[above] {$0$} (D);
		%\path [->] (A)  edge[bend right=80]  node[above] {$0$} (E);

	\end{tikzpicture}
	\caption{Reduced graph $G_{W_1}$ for $D$ (Figure \ref{ex0}), where $W_1=\{2,4\}$. }
	\label{ind1}
\end{subfigure}
\begin{subfigure}[b]{0.45\textwidth}
%\begin{figure}[h!]
	%\centering
	\begin{tikzpicture}%[scale=0.4]
		
		\node[shape=circle,draw=red,line width=1mm] (B) at (2,0) {1};

		%\node[shape=circle,draw=green,line width=1mm] (D) at (3.5,2) {2};
		\node[shape=circle,draw=green,line width=1mm] (C) at (6,2) {3};

		%\node[shape=circle,draw=black] (G) at (3.5,-2) {6};
		
		\node[shape=circle,draw=black] (H) at (4.75,-2) {5};

		\path [->] (C) edge[bend left=20]  (H);
		%\path [->] (A)  edge[bend right=60]  node[above] {$0$} (D);
		%\path [->] (A)  edge[bend right=80]  node[above] {$0$} (E);
		
		\path [->] (B)  edge[bend left=30]   (C);
		%\path [->] (B)  edge[bend left=80]  node[above] {$\beta_1$} (E);

		%\path [->] (E)  edge[bend right=40]   (G);
		
		%\path [->] (G)  edge[bend left=30]   (B);
		\path [->] (H)  edge[bend left=40]   (B);

	\end{tikzpicture}
	\caption{Reduced graph $G_{W_2}$ for $D'$ (Figure \ref{ex}), where $W_2=\{1,3,5\}$.}
	\label{ind2}
\end{subfigure}
\caption{Reduced graphs associated to examples of Figure 1.}
\end{figure}
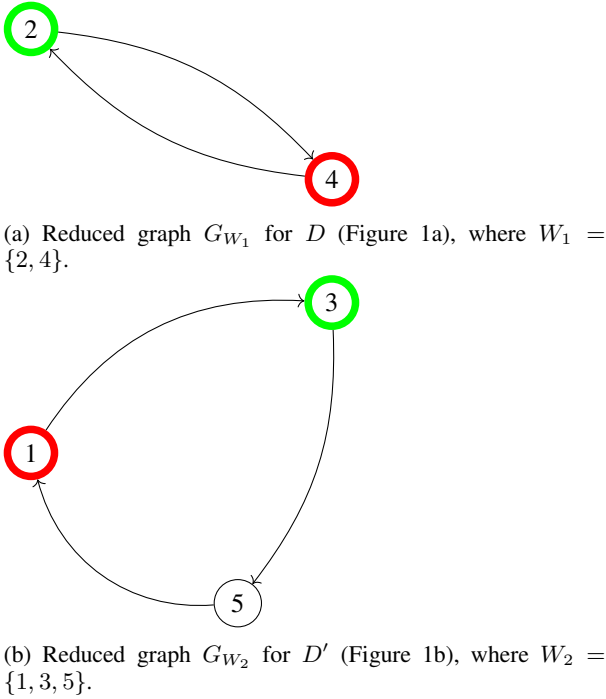

 In what follows we list the original contributions of our work.
\newline\newline\noindent 
 {\bf Statement of contribution}
\newline\newline\noindent 
 1) Differently from MP, for which a feasible solution can be detected in polynomial time \cite{feasibility}, we show that finding a feasible solution for $\C$-MP is NP-hard (see Proposition \ref{prop_CMP_hard}). We prove this by a reduction from 3-SAT.
\newline\newline\noindent 
 2) We propose a reduction method to convert $\C$-MP and $\C$-MAPF to standard MP and MAPF. This reduction is based on the partition of the vertex set $V$ into two complementary sets $W$ and $V \setminus W$, called destination and transit vertices, respectively. The elements of $W$ can be destinations for the agents, while the elements of $V \setminus W$ cannot be destinations, but can be visited by the agents while traveling between vertices in $W$.
We build a \emph{reduced graph} $G'=(W,E_W)$ where the edge set $E_W$ is defined in the following way: $(s,d) \in E_W$ if and only if there exists a directed path from $s$ to $d$ that does not violate the constraints, assuming that all vertices in $W \setminus\{s,d\}$ are occupied by an agent. We choose $W$ under the constraint that the resulting graph $G'$ is strongly connected. 
After this reduction, in order to solve $\C$-MP or $\C$-MAPF, we solve the standard MP or MAPF on $G'$. As we will see, the obtained solution can be simply converted to a solution of the original $\C$-MP or $\C$-MAPF problem.
\newline\newline\noindent 
3) It is convenient to choose a set $W$ of the largest possible cardinality.  Indeed, the larger is the cardinality of $W$, the larger is the number of agents for which we can solve MAPF on the reduced graph (see~\cite{diSC,diBOX}).
 In Theorem \ref{AC_MIS_NP}, we show that the problem of finding $W$ of maximum cardinality is strongly NP-hard. For this reason, we propose some heuristic algorithms that allow finding a suboptimal set $W$ in polynomial time.
\newline\newline\noindent 
Figures~\ref{ind1}, \ref{ind2} show possible reduced graphs $G_{W_1}$ and $G_{W_2}$ for the examples introduced above, in which $W$ has been chosen with the largest possible cardinality. In particular, the problem represented in Figure~\ref{ex0} is reduced to the graph $(W_1,E_{W_1})$ of Figure~\ref{ind1}. Since this graph contains only two nodes, we can solve a MAPF problem with only one agent, that can only move from node $2$ to $4$ and back. The motion from $2$ to $4$ corresponds to path $2-3-4$ in the original graph, while the motion from $4$ to $2$ corresponds to path $4-5-1-2$ in the original graph. The problem in Figure~\ref{ex} corresponds to the reduced graph $(W_2,E_{W_2})$ of Figure~\ref{ind2}. Since this graph has three nodes, we can solve a MAPF problem with two agents. 
It is important to remark that this reduction method has one important advantage and one important limitation. The advantage is the reduction of complexity of the resulting planning problem. Namely, while the original $\C$-MAPF and $\C$-MP are NP-hard, the reduced problems are simple MAPF and MP, and, thus, feasible solutions can be found with polynomial complexity with respect to the number of nodes and agents (\cite{diSC,WuGru09}). The disadvantage is that this reduction process is not exact. Namely, the original MAPF and MP can have feasible solutions that cannot be obtained by solving the reduced problems. This fact is evident in the reduction of $\C$-MAPF in Figure~\ref{ex0} to MAPF in Figure~\ref{ind1}. Here, the reduction allows solving MAPF with only one agent, while it is obvious that $\C$-MAPF instances with two agents can be easily solved in the original problem. Anyway, note that the fact that we lose feasible solutions is an unavoidable consequence of the complexity reduction. Obviously, if the proposed reduction were an exact one, it would preserve the computational complexity of the original problem.
\newline\noindent 
{\bf Comparison with existing literature.}
To our knowledge, $\C$-MP and $\C$-MAPF, in the formulation discussed here, have not been considered in literature. In particular, we could not find any reference that addresses our main topic: the reduction of constrained MP and MAPF problems to unconstrained ones. Anyway, some papers do consider MP and MAPF with additional constraints.
For instance,~\cite{la} studies Multi-Agent
Path Finding for Large Agents (LA-MAPF), which takes into
account the agents' occupancy, enforcing a safety
distance to avoid collisions. Differently from this work, \cite{la} does not propose a reduction of this problem into a classical MAPF, but solves it directly, using an optimal algorithm, the Multi-Constraints CBS, a generalization of the Conflict-Based Search algorithm (see~\cite{cbs}). Also, \cite{robust} introduces additional constraints to MAPF. Namely, it tightens standard collisions avoidance constraints, to avoid collisions even in case each agent experiences delays, or in the presence of localization errors. Moreover, \cite{pafp} discusses the problem of a path avoiding forbidden pairs (PAFP), which is a generalization of the classical shortest path problem. Given a set $F$ of pairs of vertices, this problem consists in finding a path for a single agent from a source to a target that constains at most one vertex from each pair in $F$.

\noindent
{\bf Notation.}
\label{sec:notation}
\noindent For a set $A$, we denote by ${\cal P}(A)$ the \emph{power set} of $A$ (i.e., the set of all subsets of $A$). A \emph{directed graph} is a pair $G=(V,E)$, where $V$ is a set of
nodes
and $E \subset \{(x,y) \in V^2 \mid x \neq y\}$ is a set of directed arcs. We focus on directed graphs, but, obviously, our results also apply to undirected graphs. Indeed, undirected graphs can be considered as a subclass of directed graphs, in which, for each edge $(u,v)\in E$, also $(v,u)\in E$.
An \textit{alphabet} $\Sigma = \{\sigma_1,\ldots,\sigma_n\}$ is a set of symbols.
A \textit{word} is any finite sequence of symbols.
The set of all words over $\Sigma$ is $\Sigma^*$, which also contains
the empty word $\epsilon$, while $\Sigma_i$ represents the set of all
words of length up to $i \in \mathbb{N}$, (i.e., words composed of up to
$i$ symbols, including $\epsilon$).
Given a word $w \in \Sigma^*$, $|w|$ denotes its length.
Given a directed graph $G = (V, E)$, we can think of $V$ as
an alphabet so that any path $p$ of $G$ is a word in $V^*$.
Namely, a \textit{path} $p$ on $G$ is a sequence of adjacent nodes of $V$
(i.e., $p= \sigma_{1}\cdots \sigma_{m}$, with $(\forall i \in \{1, \ldots, m\})\
(\sigma_i,\sigma_{i+1}) \in E$). A directed graph $G=(V,E)$ is \textit{strongly connected} if for all $v,w \in V$ there is a path from $v$ to $w$.
Given $s,t \in \Sigma^*$, the \textit{concatenation} of $s$ and $t$ is the word obtained by writing $t$ after $s$, it is denoted by $st \in \Sigma^*$;
we call $t$ a \textit{suffix} of $st$ and $s$ a \textit{prefix} of $st$.
%For $r \in V^*$, $\vec{r}$ is the rightmost symbol of $r$ (QUESTO LO USIAMO DA QUALCHE PARTE?). 
%In the following, we represent paths of $G$ as strings of symbols
%in $V$.
%This allows to use the concatenation operation on paths and to
%use prefixes and suffixes to represent portions of paths.
Given $s \in \Sigma^*$, $\textrm{Pref} (s)$ denotes the set of the prefixes of $s$. We also recall the definitions of an \textit{abstract simplicial complex} and a \textit{matroid}. 
\begin{defn}
  \label{def_ASC}
	Given a set $V$, an \textit{abstract simplicial complex} (ASC in the following) $\F$ is a family of subsets of $V$ such that the following axioms hold:\begin{enumerate}
		\item trivial axiom: $\emptyset \in \F$;
		\item hereditarian axiom: if $x \in \F$ and $y \subset x$ then $y \in \F$.
	\end{enumerate}
\end{defn}

\begin{defn}
	Given a set $V$, a \textit{matroid} $\F$ is an abstract simplicial complex such that the following axiom holds:
	\begin{itemize}
		
		\item[3.] exchange axiom: if $x,y \in F$ with $|x| = |y| + 1 $ then $ \exists i \in x-y$ such that $y \cup \{i\} \in \F$.
	\end{itemize}
      \end{defn}
      
Roughly speaking, an ASC is a family of subsets of $V$, that contains all subsets of each family element. A matroid is an ASC with the additional property that every member of the family of non-maximal cardinality is strictly included in another member.

\section{Problem Definition}
We adapt the following definitions from~\cite{diSC}.
Let $G=(V,E)$ be a digraph, with vertices $V$ and directed edges $E$. In order to comply with the terminology commonly used in literature, we will refer to agents as {\em pebbles}, but these two terms are used as synonyms (in fact, also equivalent to AGVs when thinking about the practical application).
We assign a unique label to each pebble. Set $P$ contains the pebbles labels.
A \textit{configuration} is a function $\A:P \rightarrow V$ that associates to each pebble the vertex it occupies. A configuration is \textit{valid} if it is one-to-one (i.e., each vertex is occupied by at most one pebble). Set $\chi \subset \{P \to V\}$ represents all valid configurations.
Given a configuration $\A$ and $u, v \in V$, if $u$ is occupied by a pebble and $v$ is empty, we denote by $\A[u,v]$ the configuration obtained from $\A$ moving a pebble from $u$ to $v$ without changing the positions of the other pebbles:
%\vspace{-2pt}
\begin{equation} 
	\label{c}
	\A[u,v](q):=  \Bigg\{
	\begin{array}{ll}       	
		
          v, & \text{if } \A(q)  = u ;\\
          
		\A(q), & \text{otherwise }.\\
	\end{array}
\end{equation}

Function $\rho: \chi \times E \rightarrow \chi$ is a partially defined transition function such that $\rho(\A,u \rightarrow v)$ is defined if and only if $v$ is empty and configuration $\A$ contains a pebble in $u$. In this case, $\rho(\A,u \rightarrow v)$ is the configuration obtained by moving the pebble in $u$ to $v$. Notation $\rho(\A,u \rightarrow v)!$ means that the function is well-defined. In other words,
$\rho(\A, u \rightarrow v)!$ if and only if $(u,v)\in E$, $u \in \A(P)$ and
$v \notin \A(P)$. Moreover, if $\rho(\A, u \rightarrow v)!$,
$\rho(\A, u \rightarrow v)=\A[u,v]$.

We represent plans as ordered sequences of directed edges.
It is convenient to view the elements of $E$ as the symbols of a language.
We denote by $E^*$ the Kleene star of $E$, that is the set of
ordered sequences of elements of $E$ with arbitrary length, together with the empty string $\epsilon$:
\[
E^*=\bigcup_{i=1}^\infty E^i \cup \{\epsilon\}.
\]
We extend function $\rho:\chi \times E \to \chi$ to $\rho: \chi \times E^*  \to \chi$, by setting $(\forall \A \in \chi) \rho(\A,\epsilon)!$ and $\rho(\A,\epsilon)=\A$. Moreover,  $(\forall s \in E^*, e \in E, \A \in \chi)$  $\rho(\A,  se) !$ if and only if $\rho(\A, s)!$ and $\rho(\rho(\A, s), e)!$ and, if $\rho(\A se)!$, $\rho(\A se)=\rho(\rho(\A s),e)$.
A \emph{move} is an element of $E$, and a \emph{plan} is an element of $E^*$.
Note that $\epsilon$ is the trivial plan that keeps all pebbles on their positions.
%We define an equivalence relation $\sim$ on $E^*$, by setting, for $s,t \in E^*$, $s \sim t \leftrightarrow (\forall \A \in \chi)\, \rho(\A,s)=\rho(\A,t)$.
%In other words, two plans are equivalent if they reconfigure pebbles in the same way.
\newline\newline\noindent
Before introducing the new problems, we recall the definitions of the original motion planning (MP) and multi-agent path finding (MAPF) problems.
\begin{defn}{(\textbf{MP}).}
	\label{def:mp}
	Let $G=(V,E)$ be a digraph, $P$ a set of pebbles. Given a pebble $p\in P$, an initial configuration $\A^s$, and $t\in V$, the \textit{motion planning problem} (MP) consists in finding a plan $f$ such that $t = \rho(\A^s,f)(p)$. We indicate a MP instance with $<G,s,t,\oo>$, where $s=\A^s(p)$ is the initial position of $p$ and $\oo=\A^s(P\setminus \{p\})$ is the set of initial positions of the obstacles, i.e., of all the other pebbles.
\end{defn}

\begin{defn}{(\textbf{MAPF problem}).}
\label{def:mapf}
	Given a digraph $G = (V,E)$, a pebble set $P$, an initial valid configuration $\A^s$, and a final valid configuration $\A^t$, MAPF problem $<G,\A^s,\A^t>$ consists in finding a plan $f$ such that $\A^t(P) = \rho(\A^s,f)(P)$.
\end{defn}
The only difference between MP and MAPF is the following one. In MP only one pebble is assigned a target while the others are only movable obstacles. Instead, in MAPF each pebble is assigned a final destination.

\subsection{MAPF and MP with additional constraints}
We define $\C$-MP and $\C$-MAPF by requiring that all pebbles configurations visited along a plan belong to a given family of subsets of the vertex set $V$, that represents the feasible configurations.

\begin{defn}
A constraint set is a family of subsets of $V$.
\end{defn}

%\begin{defn}
%  A family $\C$ of subsets of $V$ is an \emph{abstract simplicial complex} (ASC) if
%  \[  (\forall A,B \in \mathcal{P}(V)) A \in \C \wedge B \subset A \rightarrow B \in \C\,.
%    \]
%\end{defn}
According to Definition~\ref{def_ASC}, constraint set $\C$ is an ASC if every subset of an admissible configuration is admissible. In other words, any pebble configuration obtained by removing one or more pebbles from a feasible configuration is still feasible. 
Obviously, not all constraint sets are ASC. For instance, the constraint set defined by requiring that one subset of $V$ contains \emph {at most} one pebble is an ASC. Instead, the constraint set obtained by requiring that this subset contains \emph{exactly} one pebble is not.
Anyway, the family of constraint sets that are ASC is sufficiently broad to include many cases encountered in applications.

For instance, to define $\C$, we may choose a family ${\cal Q}$ of couples $(S,k)$ where $S \subset V$ and $k \in \mathbb{N}$.
For each couple $(S,k)$, we require that, at any time, no more than $k$ agents occupy the nodes in $S$. In this way, we define the following constraint set $\C_{\cal Q}$, associated to ${\cal Q}$.

\begin{equation}
\label{eq:cq}
\C_{\cal Q}=\{U \subset V: (\forall (S,k) \in {\cal Q}) \quad   |S \cap U| \leq k\}.
\end{equation}
%A further special case is obtained when $k=1$ in each couple, i.e.: 
%\begin{equation}
%\label{eq:cq1}
%\C=\{U \subset V: \forall S \in {\cal S} \quad   |S \cap U| \leq 1\},
%\end{equation}
%where ${\cal S}$ is a collection of subsets of $V$.
Note that (\ref{eq:cq}) corresponds to the definition of constraint given in the Introduction, which is also the most natural one for the applications with AGVs. It turns out that form~(\ref{eq:cq}) is not restrictive, that is, a constraint set is an ASC if and only if it can be written in this form.

\begin{prop}
  \label{prop_equiv}
  A constraint set $\C$ is an ASC if and only if there exists a family ${\cal Q}$ of pairs $(S,k)$ such that $\C=\C_{\cal Q}$.
\end{prop}
\begin{proof}
  $(\Leftarrow)$ First note that $\emptyset \in \C_{\cal Q}$, so that 1), of Definition~\ref{def_ASC}, holds. To show that also 2) holds, let $A \in \C_{\cal Q}$ and $B \subset A$. Then, 
  $(\forall (S,k) \in {\cal Q)}\, |A \cap S| \leq k$, and, since $B \subset A$, also
  $(\forall (S,k) \in {\cal Q})\, |B \cap S| \leq k$, which implies that $B \in \C_{\cal Q}$.
\newline\newline\noindent
$(\Rightarrow)$
  Let $\C$ be an ASC. For each $A \in \mathcal{P}(V)$, set
  $v_A=\max \{|B|,\, B \subset A,\, B \in \C\}$ and define
  the family ${\cal Q}$ composed of pairs $(A,v_A)$, for all $A \in \mathcal{P}(V)$.
  For every $A \in \mathcal{P}(V)$, we have that
  $(\forall C \in \C)\, |C \cap A| \leq v_A$. This implies that $\C \subseteq \C_{\cal Q}$. 
  To prove that also $\C_{\cal Q} \subseteq \C$, by contradiction,
  assume that there exists $M \in \C$ such that $M \notin \C_{\cal Q}$. Then, there exists $N \in \mathcal{P}(V)$ such that $|M \cap N|>v_N$. However, since $\C$ is an ASC and $M \in \C$, necessarily $M \cap N \in \C$, which implies that $v_N \geq |M \cap N|$.
\end{proof}

Throughout this paper, we make the following assumption.

\begin{assum}
The constraint set $\C$ is an ASC or, equivalently, $\C=\C_{\cal Q}$ for some  family ${\cal Q}$ of pairs $(S,k)$.
\end{assum}
%Obviously, not all constraint sets are ASC. For instance, the constraint set defined by requiring that one subset of $V$ contains \emph {at most} one pebble is an ASC. Instead, the constraint set obtained by requiring that this subset contains \emph{exactly} one pebble is not.
%Anyway, the family of constraint sets that are ASC is sufficiently broad to include many cases encountered in applications.

As a consequence of Proposition~\ref{prop_equiv}, we will usually define the constraint set $\C$ by choosing a family $\cal Q$ of pairs $(S,k)$, with the understanding that the actual constraint set is $\C_{\cal Q}$.

\begin{rem}
An equivalent alternative to the definition of a constraint set through a  family $\cal Q$ of pairs $(S,k)$ is its definition through a family  $\cal Q'$ of triples $(S,k,w^S)$, where $w^S$ is a vector of positive integer weights.
In this case the constraint set is defined as follows
\begin{equation}
\label{eq:cq1}
\C_{\cal Q'}=\{U \subset V: (\forall (S,k,w^S) \in {\cal Q'}) \quad   \sum_{i\in S \cap U} w_i^S \leq k\}.
\end{equation}
Basically, each constraint can be viewed as a {\em knapsack constraint}, where $w^S$ is the weight vector of the objects in $S$ and $k$ is the weight capacity of the knapsack. 
Definitions (\ref{eq:cq}) and (\ref{eq:cq1}) are equivalent. It is obviously true that each constraint set defined as in (\ref{eq:cq}) can be represented as a constraint set defined as in (\ref{eq:cq1}) (just set $w_i^S=1$ for all $i\in S$ and for all $S$).
But also the opposite is true. Indeed, each knapsack constraint $\sum_{i\in S \cap U} w_i^S \leq k$ can be replaced by the collection of all its cover inequalities (see, e.g., \cite{Balas75}), i.e.,
$$
\forall C \subseteq S: \ \sum_{i\in C} w_i^S > k\ \ \rightarrow\ \ |C \cap U| \leq |C|-1,
$$
i.e., each triple $(S,k,w^S)$ can be replaced by the collection of pairs $(C, |C|-1)$ for all $C\subseteq S$ such that $\sum_{i\in C} w_i^S > k$.
Note that the collection of cover inequalities associated to a knapsack constraint may contain an exponential number of elements. Thus, while equivalent, representation (\ref{eq:cq1}) is usually more compact than representation  (\ref{eq:cq}).
\end{rem}

%\begin{defn}
%  \label{def_con}
%	A subset of nodes $U \subset V$ is \textbf{consistent with constraints} $\C$ if 
%	
%	\[\forall (S,k) \in \C \quad   |S \cap U| \leq k\]
%
%      \end{defn}

%We call $V_{\C} \subset \mathcal{P}(V)$ the family of subsets of $V$ that are \%textbf{consistent with constraints} $\C$.
        
A couple $(f,\A^s)$, where $f \in E^*$ is a plan and $\A^s: P \to V$ is a valid initial configuration, is \textbf{consistent with} $\C$ if every pebble configuration visited along the plan belongs to $\C$, that is
%QUI SECONDO ME VA SCRITTA MEGLIO LA DEFINIZIONE DI CONSISTENT E VA FATTO RIFERIMENTO A UNA QUALCHE CONFIGURAZIONE INIZIALE $\A$.
        \[
(\forall s \in \textrm{Pref} (f))  \quad \rho(s,\A^s) \in \C.
\]
Note that this definition implicitly assumes that $\rho(s,\A^s)$ is well-defined for every prefix $s$ of $f$. That is, all moves of $f$ move a pebble to an empty node.

%        such that $\A^t = \rho(f,\A^s)$, we say that $f$ is  if, denoting with  $\A^0:=\A^s$, $\A^{i+1} := \rho(\A^i, e_i)$ and  $\A^n:=\A^t$ the intermediate configurations, it holds that
	
%	\[ \forall i=0,...,n \quad \A^{i}(P) \text{ is consistent with the constraints } \C. \]

%\end{defn}
%In the AC-MP and AC-MAPF problems we require that all pebble configurations, including the intermediate ones, belong to $\C$.
In the following, we will consider MP and MAPF with the additional requirement that the visited pebble configurations are consistent with a constraint set $\C$.
In order to highlight the dependency on $\C$, we will denote such problems by  $\C$-MP and $\C$-MAPF. Their definitions, which are simple extensions of Definitions \ref{def:mp} and \ref{def:mapf}, are the following.

\begin{defn}{(\textbf{$\C$-MP}).}
	\label{def:mpc}
	Let $G=(V,E)$ be a digraph, $P$ a set of pebbles, and $\C$ a constraint set. Set  $p\in P$, $s,t \in V$, $\oo \subset V$.
        Let $\A^s$ be an initial configuration such that $\A^s(p)=s$ and $\A^s(P\setminus \{p\})=\oo$.
        The $\C$-MP defined by $<G,(s,t),\oo,\C>$ consists in finding a plan $f$ such that 
	\begin{enumerate}
		\item $t = \rho(\A^s,f)(p)$;
		
		\item $(f,\A^s)$ is consistent with $\C$.
	\end{enumerate} 
\end{defn}

In Definition~\ref{def:mpc}, $s$ and $t$ are the source and target vertices for pebble $p$, respectively. Set $\oo$ contains the initial positions of all other pebbles.

\begin{defn}{(\textbf{$\C$-MAPF problem}).}
	Let $G = (V,E)$ be a digraph, $P$ a pebble set, and $\C$ a constraint set. Given an initial valid configuration $\A^s$, and a final valid configuration $\A^t$, the $\C$-MAPF problem defined by $<G,\A^s,\A^t, \C>$ consists in finding a plan $f$ such that 
	
	\begin{enumerate}
		\item $\A^t(P) = \rho(\A^s,f)(P)$;
		\item $(f,\A^s)$ is consistent with $\C$.
	\end{enumerate}
\end{defn}

\section{Complexity of the motion planning problem with additional constraints}
In \cite{GMP1R}, it is proved that finding a solution for an MP instance with a minimum number of moves is NP-hard. Instead, in \cite{feasibility} it is shown that the feasibility of MP on an acyclic digraph $G=(V,E)$ can be decided in linear time with respect to $|V|\cdot|E|$, while reference \cite{WuGru09} generalizes this result showing that on directed graphs it can be decided in time $O(|V|^2|E|)$. 

In this section we will prove that, differently from MP, even establishing the feasibility of $\C$-MP is NP-hard.

\begin{figure}[h!]
	\centering
	 \resizebox{\columnwidth}{!}{
	\begin{tikzpicture}
		\node[shape=circle,draw=green,line width=1mm] (A) at (0,0) {0};
		\node[shape=circle,draw=black] (B) at (2,0) {1};
		\node[shape=circle,draw=black] (C) at (5,0) {3};
		\node[above] at (5,0.2) {$\bar{x}_2$};
		\node[shape=circle,draw=black] (D) at (5,2) {2};
		\node[above] at (5,2.2) {$x_1$};
		\node[shape=circle,draw=black] (E) at (8,0) {4};
		\node[below] at (7.9,-0.25) {$x_3$};
		\node[shape=circle,draw=black] (F) at (11,0) {6};
		\node[above] at (11,0.2) {$x_2$};
		\node[shape=circle,draw=black] (G) at (11,-2) {7};
		\node[above] at (11,-1.8) {$\bar{x}_3$};
		\node[shape=circle,draw=black] (H) at (11,2) {5};
		\node[above] at (11,2.2) {$\bar{x}_1$};
		\node[shape=circle,draw=blue,line width=1mm] (I) at (14,0) {8} ;
		
		\node[shape=circle,draw=red,line width=1mm] (J) at (1,5) {9} ;
		\node[shape=circle,draw=black] (K) at (3,4) {10} ;
		\node[above] at (3.1,4.3) {$\bar{x}_1$};
		\node[shape=circle,draw=black] (L) at (3,6) {11} ;
		\node[above] at (3.1,6.3) {$x_1$};
		
		\node[shape=circle,draw=red,line width=1mm] (M) at (6,5) {12} ;
		\node[shape=circle,draw=black] (N) at (8,4) {13} ;
		\node[above] at (8.1,4.3) {$\bar{x}_2$};
		\node[shape=circle,draw=black] (O) at (8,6) {14} ;
		\node[above] at (8.1,6.3) {$x_2$};
		
		\node[shape=circle,draw=red,line width=1mm] (P) at (11,5) {15} ;
		\node[shape=circle,draw=black] (Q) at (13,4) {16} ;
		\node[above] at (13.1,4.3) {$\bar{x}_3$};
		\node[shape=circle,draw=black] (R) at (13,6) {17} ;
		\node[above] at (13.1,6.3) {$x_3$};
		
		\path [->] (A) edge  (B);
		\path [->] (D) edge[bend left=40]  (E);
		%\path [->] (A)  edge[bend right=60]  node[above] {$0$} (D);
		%\path [->] (A)  edge[bend right=80]  node[above] {$0$} (E);
		\path [->] (B) edge  (C);
		\path [->] (B)  edge[bend left=40]   (D);
		%\path [->] (B)  edge[bend left=80]  node[above] {$\beta_1$} (E);

		\path [->] (C) edge  (E);
		\path [->] (F) edge  (I);
		\path [->] (E)  edge[bend left=40]   (H);
		\path [->] (E)  edge[bend right=40]   (G);
		\path [->] (E) edge  (F);
		\path [->] (G)  edge[bend right=40]   (I);
		\path [->] (H)  edge[bend left=40]   (I);
		
		\path [->] (J) edge  (K);
		\path [->] (J) edge  (L);
		\path [->] (M) edge  (N);
		\path [->] (M) edge  (O);
		\path [->] (P) edge  (Q);
		\path [->] (P) edge  (R);
	\end{tikzpicture}
}
	\caption{Reduction of the 3-SAT instance $(x_1\vee \bar{x}_2) \wedge x_3 \wedge (\bar{x}_1\vee x_2 \vee \bar{x}_3)$ to an $\C$-MP instance.}
	\label{complexity_ex}
\end{figure}
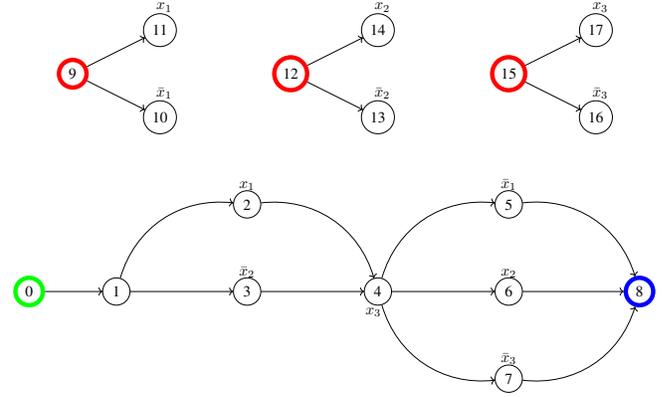

The proof of NP-hardness is by a reduction
from 3-SAT. In particular, it uses a construction similar to that one used in \cite{pafp} to prove NP-hardness of PAFP problem, even if the two problem are different.
As an example, we consider the 3-SAT problem with variables $V=\{x_1,x_2,x_3\}$ and conjunctive normal form $(x_1\vee \bar{x}_2) \wedge x_3 \wedge (\bar{x}_1\vee x_2 \vee \bar{x}_3)$. This problem can be reduced to a $\C$-MP instance $<G,(s,t),\mathcal{O},\C>$, where $G$ is the digraph displayed in Figure~\ref{complexity_ex}, $(s,t)=(0,8)$, $\oo = \{9,12,15\}$, and $\C$ is defined as in (\ref{eq:cq}) where, for each pair $(S,k)\in {\cal Q}$, $k=1$ and $S$ belongs to the collection: 
\begin{gather*}
\mathcal{S}=\left\{\{11,5\}, \{10,2\}, \{3,14\}, \{6,13\}, \{4,16\},\right.\\ \left.\{7,17\}, \{1,9\}, \{1,12\}, \{1,15\}\right\}.
\end{gather*}
Indeed, solving $<G,(0,8),\{9,12,15\},\C>$ means moving the obstacles so that there is a feasible path for a pebble $p$ from $0$ to $8$. Since $p$ must necessarily pass through $1$, the 3 obstacles must leave from their initial positions (since $\{1,9\},\{1,12\}, \{1,15\}\in \mathcal{S}$). 
Therefore, they must move so that at least one between $\{2,3\}$ ($\{x_1,\bar{x}_2\}$), at least one between $\{5,6,7\}$ ($\{\bar{x}_1,x_2,\bar{x}_3\}$), and $4$ ($\{x_3\}$) remains free, which is equivalent to solve  the 3-SAT associated to form $ (x_1\vee \bar{x}_2) \wedge x_3 \wedge (\bar{x}_1\vee x_2 \vee \bar{x}_3)$.

The above procedure can be extended to reduce any 3-SAT instance to an $\C$-MP instance on a digraph, so that we can prove the following.
\begin{prop}
  \label{prop_CMP_hard}
	$\C$-MP is NP-hard.
\end{prop}

\begin{proof}
Let us consider a 3-SAT instance, consisting of variables $x_1,\ldots,x_n$ and
$k$ clauses, with at most 3 literals each. Formally, the $j$-th clause is $C_j=c^{1}_j \vee c^{2}_j \vee c^{3}_j$, $j\in \{1,\ldots,k\}$, with $c_j^m = x_i$ or $c_j^m = \bar{x}_i$, $m\in \{1,2,3\}$, for some $i\in \{1,\ldots,n\}$.  We define a $\C$-MP instance on a digraph $G$ as follows. Let $\oo = \{o_1,o_2,...,o_{n}\}$ be the set of nodes that correspond to initial obstacles positions. For $i=1,\ldots,n$, we define a weakly-connected component of graph $G$, which connects node $o_i$ to two nodes, representing literals $x_i$ and $\bar{x}_i$ (see Figure~\ref{c1}).
Moreover, we add to $G$ a weakly-connected component, which represents the sequence of $k$ clauses.
This component contains nodes $s,t,d_1,\ldots,d_k$ and
$c_j^m,$ $j=1,\ldots,k$, $m=1,2,3$ (see Figure~\ref{c2}).
Note that $c_j^m$ represents both the literals and some node of the graph.

Nodes $s$ and $t$, are the source and the target of the marked pebble. Node $c_k^i$ is associated to the literal containing variable $x_i$ or $\bar{x}_i$ in the $k$-th clause.  Nodes $d_1,\ldots,d_k$ connect the subcomponents $C_1,\ldots,C_k$, associated to each clause.
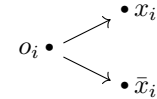
\begin{figure}[h!]
\centering
	\begin{tikzpicture}
		\node (A) at (0,0) {\textbullet};
		\node[left] at (0,0) {$o_i$};
		\node (B) at (1,0.5) {\textbullet};
		\node[right] at (B) {$x_i$};
		\node (C) at (1,-0.5) {\textbullet};
		\node[right] at (C) {$\bar{x}_i$};

		\draw[->] (A) -- (B);
		
		\draw[->] (A) -- (C);

		%\path [->] (A)  edge[bend right=60]  node[above] {$0$} (D);
		%\path [->] (A)  edge[bend right=80]  node[above] {$0$} (E);

	\end{tikzpicture}
%	\label{s}
%\end{figure}
%Moreover, we construct another component of the digraph $G$ which represents the sequence of $k$ clauses, which connects nodes $s$ to node $t$, respectively the source and the target of the pebble.
%	\begin{figure}[h!]
%	\centering
\caption{Component $\{o_i, x_i,\bar{x}_i\}$.}
	\label{c1}
\end{figure}
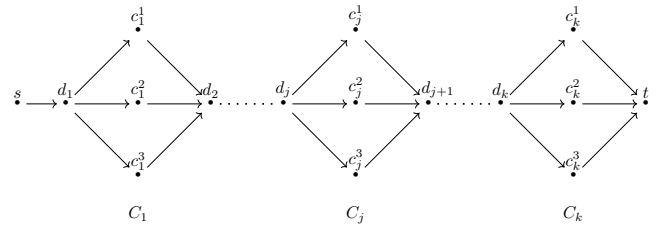
\begin{figure}[h!]
	\resizebox{\columnwidth}{!}{
\centering
\begin{tikzpicture}
		\node (S) at (-1,0) {\textbullet};
		
		\node[above] at (S) {$s$};
		\node (A) at (0,0) {\textbullet};
		\node[above] at (0,0) {$d_1$};
		\node (B) at (1.5,1.5) {\textbullet};
		\node[above] at (B) {$c_1^1$};
		\node (C) at (1.5,-1.5) {\textbullet};
		\node[above] at (C) {$c_1^3$};
			\node[below] at (1.5,-2) {$C_1$};
		\node (D) at (1.5,0) {\textbullet};
		\node[above] at (D) {$c_1^2$};
		\node (E) at (3,0) {\textbullet};
		\node[above] at (E) {$d_2$};

		\node (I) at (4.5,0) {\textbullet};
		\node[above] at (I) {$d_j$};

		\node (F) at (6,1.5) {\textbullet};
		\node[above] at (F) {$c_j^1$};
			\node[below] at (6,-2) {$C_j$};
		\node (G) at (6,-1.5) {\textbullet};
		\node[above] at (G) {$c_j^3$};		
		\node (H) at (6,0) {\textbullet};
		\node[above] at (H) {$c_j^2$};

		\node (J) at (7.5,0) {\textbullet};
		\node[above] at (7.7,0) {$d_{j+1}$};

		\node (K) at (9,0) {\textbullet};
		\node[above] at (K) {$d_k$};
		
			\node[below] at (10.5,-2) {$C_k$};
		\node (L) at (10.5,1.5) {\textbullet};
		\node[above] at (L) {$c_k^1$};
		\node (M) at (10.5,-1.5) {\textbullet};
		\node[above] at (M) {$c_k^3$};		
		\node (N) at (10.5,0) {\textbullet};
		\node[above] at (N) {$c_k^2$};
			\node (T) at (12,0) {\textbullet};
		\node[above] at (T) {$t$};
		
	\draw[->] (S) -- (A);	
		\draw[->] (A) -- (B);
		
		\draw[->] (A) -- (C);
			\draw[->] (A) -- (D);
			\draw[->] (D) -- (E);
			\draw[->] (B) -- (E);
			\draw[->] (C) -- (E);
		
		\draw [loosely dotted, line width=1pt, line join=round] (E) -- ++(I);
		
			\draw[->] (I) -- (F);
				\draw[->] (I) -- (G);
					\draw[->] (I) -- (H);

					\draw[->] (F) -- (J);
					\draw[->] (G) -- (J);
					\draw[->] (H) -- (J);
	\draw [loosely dotted, line width=1pt, line join=round] (J) -- ++(K);	
		\draw[->] (K) -- (L);
	\draw[->] (K) -- (M);
	\draw[->] (K) -- (N);
	
		\draw[->] (L) -- (T);
	\draw[->] (M) -- (T);
	\draw[->] (N) -- (T);
		%\path [->] (A)  edge[bend right=60]  node[above] {$0$} (D);
		%\path [->] (A)  edge[bend right=80]  node[above] {$0$} (E);

	\end{tikzpicture}}
\caption{Component associated to the $k$ clauses.}
	\label{c2}
\end{figure}
We define the constraint set $\C$ in form (\ref{eq:cq}), where, for each pair $(S,k)\in {\cal Q}$, $k=1$ and $S$ is the union of two families of sets.
The first family is:
\[\{\{d_1,o_i\}, i= 1,...,n \}.
\]
Since the pebble must pass through $d_1$ to reach target $t$, this constraint set forces the obstacles to move away from their initial positions and to choose one of the two nodes $x_i$ or $\bar{x}_i$. The second family is:
%For each variable, this choice depends on the second collection
% of sets: 
$$
 \left\{ 
\begin{array}{ll}
\{c_j^m,x_i\} & \mbox{if}\ c_j^m=x_i, \\ [6pt]
\{c_j^m,\bar{x}_i\} & \mbox{if}\ c_j^m=\bar{x}_i,
\end{array} j= 1,\ldots,k,\;  m=1,2,3
\right\}.
$$
%\[ \forall j= 1,...,k \;\;\; \forall m=1,2,3  \; \; \;\text{if} \; \;c_j^m=x_i \; \Rightarrow \;\; \{c_j^m,\bar{x}_i\} ,\]
%\[ \quad \quad \quad \quad \quad \quad \quad \quad \quad \quad \quad \quad \text{if} \; \;c_j^m=\bar{x}_i \; \Rightarrow \;\; \{c_j^m,x_i\}. \]
In this way, an obstacle placed at node $x_i$ disables all nodes associated to literal $x_i$, that is, belonging to set 
$\{c_j^m=x_i, j=1,\ldots,k, m=1,2,3\}$ .
The marked pebble is able to reach node $t$ if and only if the obstacles are placed in such a way that each subcomponent $C_j$ contains at least one node that has not been disabled. This is equivalent to find an assignment of truth values to the variables $x_i$ that makes all clauses true in the original 3-SAT problem.
Note that, if the $\C$-MP problem is feasible, the negated literal associated to the nodes of the final obstacles positions are a feasible solution of the 3-SAT problem.
\\
\end{proof}

Since $\C$-MAPF is more general than $\C$-MP (that is, every $\C$-MP instance can be solved by solving a $C$-MAPF instance, in which the final positions of the unmarked pebbles are arbitrarily assigned) the following is a direct consequence of Proposition~\ref{prop_CMP_hard}.

\begin{prop}
$\C$-MAPF is NP-hard.
  \end{prop}

\section{Reduction of $\C$-MAPF into MAPF}
Our main idea for solving $\C$-MAPF (and, similarly, $\C$-MP) is to transform a $\C$-MAPF instance $<G,\A^s,\A^t,\C>$ into a MAPF instance $<G',\A^s,\A^t>$. To do that, we convert the $\C$-MAPF problem on the original graph into a MAPF problem on a new graph (the \emph{reduced graph}), in general with a lower number of vertices, where the constraints can be ignored (in the sense that they are satisfied by construction).
A solution of the classical MAPF (or MP) on the reduced graph can be converted to a solution of $\C$-MAPF (or $\C$-MP) on the original graph.\\

% The reduced graph $G'=(V',E')$ should be such that $V'$ is a subset of $V$, and the constraints restricted to $V'$ are always satisfied. Moreover, we want $G'$ to be as large as possible, so that as many pebbles as possible can move on it.

\noindent In the following, for a vertex set $V$, an ASC $\C$, and $Z \subset V$, we define $V^{\C}_Z=\{x \in V: Z \cup \{x\} \in \C\}$.

\begin{lem}
	\label{lemma_V_cont}
	Let $G=(V,E)$ be a digraph, $\C$ an ASC, and let $Z \subset W \subset V$.
	Then, $V^{\C}_Z \supset V^{\C}_W$.
\end{lem}

\begin{proof}
	
	Set $x \in V^{\C}_W$, then $Z \subset W$ implies that $\C \ni W \cup  \{x\} \supset Z \cup \{x\}$. Since $\C$ is an ASC, this implies that $Z \cup \{x\} \in \C$.
\end{proof}

The reduced graph associated to a vertex subset $W \subset V$ is defined as follows.

\begin{defn}
\label{def:redgraph}
Let $G=(V,E)$ be a digraph, $\C$ a constraint set, and $W \subset V$ a subset of nodes with $W \in \C$.
The reduced digraph $G_W=(W,E_W)$ is such that
\begin{gather*}
\forall v_1,v_2 \in W \; (v_1,v_2)\in E_W\;  \text{if and only if } \\ \text{there exists a directed path on } G_W^{v_1v_2} \text{ from } v_1 \text{ to } v_2,
\end{gather*}
where $G_W^{v_1 v_2}=(V_{W}^{v_1,v_2},E_{W}^{v_1,v_2})$ is obtained from $G$ erasing:

%\begin{enumerate}
	%\item $\not \exists S \in \C$ such that $v_1,v_2 \in S$,
	%\item it exists a path on $G_W^{v_1v_2}$ from $v_1$ to $v_2$,
%\end{enumerate} 

\begin{itemize}
	\item the vertices in $W\setminus \{v_1,v_2\}$;
	%\item all the vertices of $S \setminus \{v_1,v_2\}$ with $S \in \C$ such that $S \cap ( W\setminus \{v_1,v_2\}) \not = \emptyset$.
\item all the vertices $v\in V\setminus W$ such that $\{v\}\cup [W\setminus \{v_1,v_2\}]\not \in {\cal C}$ (i.e., $v \not \in V^{\C}_{W\setminus \{v_1,v_2\}}$).
\end{itemize} 	
\end{defn}

In other words, $G_{W}^{v_1 v_2}$ is the subgraph of $G$ obtained by removing all nodes in $W$ (apart from $v_1$ and $v_2$) and all nodes $v \in V$, such that, the subset obtained by adding $v$ to $W \setminus \{v_1,v_2\}$ does not belong to constraint set $\C$. This definition is justified by the fact that a pebble placed in $v_1$ can move to $v_2$ on subgraph $G_W^{v_1 v_2}$ without violating the constraints, even if all nodes in $W \setminus \{v_1,v_2\}$ are occupied. Indeed, $G_W^{v_1 v_2}$ contains only those nodes that can be safely added to $W \setminus \{v_1,v_2\}$ without violating the constraints.
%Therefore, if $G_W^{v_1 v_2}$ contains a directed path from $v_1$ to $v_2$, then edge $(v_1,v_2)$ is present in $E_W$.

\begin{defn}
	\label{family}
	Let $G=(V,E)$ be a digraph and let $\C$ be a constraint set. A non-empty subset $W \subset V$ is \textbf{independent} on $(G,\C)$ if $G_W$ is strongly connected. We denote by $\F_G^{\C}$ the family of all independent subsets of $V$. The empty subset $\emptyset$ is always independent on $(G,\C)$.
\end{defn}

\begin{figure}[h!]
	\centering
	\begin{tikzpicture}%[scale=0.4]

		\node[shape=circle,draw=green,line width=1mm] (D) at (3.5,2) {2};
		
		\node[shape=circle,draw=red,line width=1mm] (E) at (7.5,0) {4};
		
		%\node[shape=circle,draw=black] (G) at (3.5,-2) {6};
		
		\node[shape=circle,draw=black] (H) at (4.75,-2) {5};

		\path [->] (D) edge[bend left=20]  (H);
		
		\path [->] (D) edge[bend left=20]  (E);
		%\path [->] (A)  edge[bend right=60]  node[above] {$0$} (D);
		%\path [->] (A)  edge[bend right=80]  node[above] {$0$} (E);

		%\path [->] (B)  edge[bend left=80]  node[above] {$\beta_1$} (E);

		\path [->] (E)  edge[bend left=30]   (H);
		%\path [->] (E)  edge[bend right=40]   (G);
		%\path [->] (H)  edge[bend left=20]   (G);	

	\end{tikzpicture}
	\caption{Reduced graph $G_{W_3}$ for $D'$ (Figure \ref{ex}), where $W_3 =\{2,4,5\}$ is not independent. }
	\label{ind3}
\end{figure}
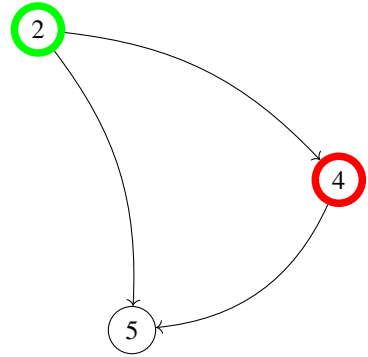

For instance, consider graphs $D$ and $D'$ (and the corresponding constraint sets $\C$, $\C'$) in Figures~\ref{ex0} and~\ref{ex}, respectively. Subset $W_1=\{2,4\}$ is independent both on $(D,\C)$ and on $(D',\C')$ (see Figure \ref{ind1}), while subset $W_2 =\{1,3,5\}$ is independent only on $(D',\C')$ (see Figure \ref{ind2}). Moreover, subset $W_3=\{2,4,5\}$ is independent neither on $(D,\C)$ nor on $(D',\C')$. In particular, on $D'$, $(5,2)\not\in E_{W_3}$, since on $D'$ there does not exist a path from $5$ to $2$, after the removal of node $\{4\}=\{2,4,5\}\setminus \{2,5\}$ and of node $1$, since $\{1,4\} \notin \C$ (see Figure \ref{ind3}).

In what follows, we show that the family $\F_G^{\C}$ of all independent subsets of $V$ is an ASC. Given a digraph $G$, the contraction of a vertex $v$ is the digraph obtained from $G$ by eliminating $v$ and merging all incoming and outgoing edges of $v$.

\begin{defn}
  Let $G=(V,E)$ be a digraph and $v \in V$. The graph obtained from $G$ by contracting vertex $v$ is defined as $G/v=(V\setminus\{v\}, E')$, where $E'$ contains

  i) all $e \in E$ such that $e \cap\{v\}= \emptyset$,

  ii) all $(u,w)$ such that $\{(u,v)(v,w)\} \subset E$.
\end{defn}

For completeness, we state the following well-known fact.

\begin{lem}
  \label{lem_str_con}
Let $G=(V,E)$ be a strongly connected digraph and $v \in V$. Then, $G/v$ is strongly connected.
\end{lem}

\begin{proof}
  We need to show that there exists a directed path in $G/v$ between any couple of vertices $u,w$.
Since $V$ is strongly connected, there is a directed path from $u$ to $w$ on $G$. If this path does not contain $v$, then this is also a path in $G/v$. If this path contains $v$, it can be written as $p=p_1 v^- v v^+ p_2$, where $v^-$ and $v^+$ are the predecessor and the successor of $v$ in this path. By definition, $G/v$ contains edge $(v^-,v^+)$ so that $p_1 v^- v^+ p_2$ is a path on $G/v$.  
\end{proof}

\begin{prop}
  \label{prop_subgraph}
  Let $G=(V,E)$ be a strongly connected digraph, let $W \subset V$, $w \in W$. Then, $G_{W \setminus \{w\}} \supset G_W/w$.
\end{prop}

\begin{proof}
  Let $(u,v)$ be an edge of $G_W/w$ and let $G_W=(W,E_W)$. By definition of $G_W/w$, one of the following holds:
  \begin{itemize}
  \item[i)] $(u,v) \in  E_W$,

  \item[ii)] $(u,w),(w,v) \in E_W$.
   
   \end{itemize}
  \noindent In case i), since $(u,v) \in E_W$, $G_W$ contains a directed path $p$ from $u$ to $v$ that belongs to $V^{\C}_{W\setminus\{u,v\}}$.
Set $Z=W \setminus \{w\}$. Since $Z \subset W$, by Lemma~\ref{lemma_V_cont}, $V^{\C}_{Z\setminus\{u,v\}} \supset V^{\C}_{W\setminus\{u,v\}}$. Hence, $p$ belongs  to $V^{\C}_{Z\setminus\{u,v\}}$, and $(u,v)$ is a directed edge of $G_Z$.
In case ii), $G$ contains a directed path $p_1$ from $u$ to $w$ that belongs to  $V^{\C}_{W\setminus\{u,w\}}$, and a directed path $p_2$ from $w$ to $z$ that belongs to  $V^{\C}_{W\setminus\{w,z\}}$. Since $Z=W \setminus \{w\}$, both paths belong to $V^{\C}_{Z\setminus\{u,v\}}$, so that their concatenation $p=p_1 p_2$ is a path from $u$ to $v$ that belongs to $V^{\C}_{Z\setminus\{u,v\}}$. This implies that $(u,v)$ is a directed edge of $G_Z$.
\end{proof}

\begin{prop}
\label{prop_Z_inde}
Let $G=(V,E)$ be a strongly connected digraph and $\C$ a constraint set. Let $W \subset V$ be independent on $(G,\C)$, and let $Z \subset W$. Then, $Z$ is independent on $(G,\C)$.

\end{prop}

\begin{proof}

  Let $\{v_1,v_2,\ldots,v_m\}=W \setminus Z$. Since $W$ is independent on $(G,\C)$,  $G_W$ is strongly connected. By Lemma~\ref{lem_str_con}, $G_W/v_1$ is also strongly connected.
  By Proposition~\ref{prop_subgraph}, $G_W/v_1$ is a subgraph of $G_{W \setminus \{v_1\}}$, which implies that $G_{W \setminus \{v_1\}}$ is also strongly connected.
  By reiterating the same argument, we obtain that $G_{W \setminus \{v_1,\ldots,v_m\}}$ is strongly connected. Hence, $Z$ is independent on $(G,\C)$.
\end{proof}

\begin{prop}
\label{prop:fgc}
Let $G=(V,E)$ be a strongly connected digraph, and let $\F_G^{\C}$ be a family of subsets of $V$, defined as in Definition~\ref{family}. Then, $\F_G^{\C}$ is an ASC.
% BIOSGNEREBBE DIRE CHE \'E SODDISFATTO IL TRIVIAL AXIOM OVVERO DIRE CHE IL GRAFO VUOTO VIENE CONSIDERATO COME FORTEMENTE CONNESSO. 
\end{prop}

\begin{proof}  
In Definition~\ref{def_ASC}, 1) is satisfied since, by definition of $\F_G^{\C}$, $\emptyset \in \F_G^{\C}$ , 2) is a direct consequence of Proposition~\ref{prop_Z_inde}.
\end{proof}

\begin{obser}
In the general case, $\F_G^{\C}$ is not a matroid. For instance, consider the following counterexample.
	
	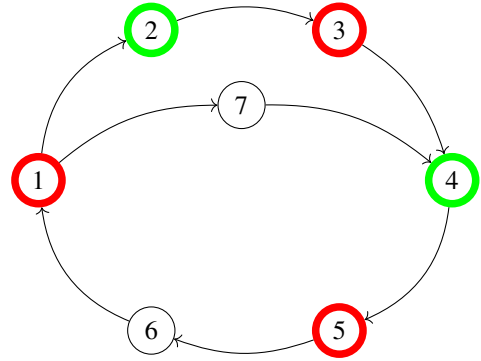
\begin{figure}[h!]
		\centering
		\begin{tikzpicture}
			
			\node[shape=circle,draw=red,line width=1mm] (B) at (2,0) {1};
			\node[shape=circle,draw=black] (I) at (4.7,1) {7};
			\node[shape=circle,draw=red,line width=1mm] (C) at (6,2) {3};
			
			\node[shape=circle,draw=green,line width=1mm] (D) at (3.5,2) {2};
			
			\node[shape=circle,draw=green,line width=1mm] (E) at (7.5,0) {4};
		
			\node[shape=circle,draw=black] (G) at (3.5,-2) {6};
			
			\node[shape=circle,draw=red,line width=1mm] (H) at (6,-2) {5};

			\path [->] (B) edge[bend left=20] (I);
			\path [->] (I) edge[bend left=20] (E);
			\path [->] (D) edge[bend left=20]  (C);
			%\path [->] (A)  edge[bend right=60]  node[above] {$0$} (D);
			%\path [->] (A)  edge[bend right=80]  node[above] {$0$} (E);
			
			\path [->] (B)  edge[bend left=30]   (D);
			%\path [->] (B)  edge[bend left=80]  node[above] {$\beta_1$} (E);

			\path [->] (C) edge[bend left=20]  (E);
		
			\path [->] (E)  edge[bend left=30]   (H);
			%\path [->] (E)  edge[bend right=40]   (G);
		
			\path [->] (G)  edge[bend left=30]   (B);
			\path [->] (H)  edge[bend left=20]   (G);

		\end{tikzpicture}
		\caption{Counterexample}
		\label{counterexample}
	\end{figure}
Let $G$ be the digraph in Figure \ref{counterexample} with constraint set $\C$ defined as in (\ref{eq:cq}) where, for each pair $(S,k)\in {\cal Q}$, $k=1$ and $S$ belongs to the collection: 
$$
\mathcal{S}=\{ \{1,2\}, \{2,3\}, \{3,4\}, \{4,5\}, \{5,6\}, \{ 6,1\}\}. 
$$
Let $W=\{1,3,5\} \in \F_G^{\C}$, $U=\{2,4\} \in \F_G^{\C}$ be subsets of $V$ such that $|W|  = |U| +1$. However, $\nexists v \in W \setminus U$ such that $U \cup \{v\} \in \F_G^{\C}$. Indeed, $\forall v \in W \setminus U$ $\exists u \in U, \; S \in \mathcal{S}$ such that $u,v \in S$.  This means that the exchange axiom does not hold.
\end{obser}

\begin{prop}
  \label{prop_lift}
  Let $G=(V,E)$ be a digraph, $P$ a set of pebbles, $\C$ a constraint set, and let $W \subset V$ be independent on $(G,\C)$. Let $G_W=(W,E_W)$ be the reduced graph. For any $e=(u,v) \in E_W$, and any pebble configuration $A=\A(P) \subset W$, with $u \in A$ and $v \notin A$, there exists a directed path $p$ in $G$, from $u$ to $v$, such that
  $(\forall s \in \textrm{Pref}\, (p))\, \rho(\A,s) \in \C$.
%Moreover, if $\rho$ is the transition function for $G$ and $\rho_W$ is the transition function for $W$, then $\rho_W(\A,e)=\rho(\A,p)$.
We call such path $p$ a \textbf{\emph{lift}} of edge $e$. 
\end{prop}

\begin{proof}
  By definition of an independent set, graph $G_{W}$ is strongly connected. Since $A \setminus \{u\}  \subset W$ and since the family of independent sets $\F_G^{\C}$ is an ASC (see Proposition \ref{prop:fgc}), then also graph $G_{A \setminus \{u\}}$ is strongly connected. Hence, there is a directed path from $u$ to $v$ that visits only nodes that belong to set $\{x \in V: \{x\} \cup (A \setminus \{u\}) \in \C\}$, which implies the thesis.
\end{proof}

\begin{defn}
Let $G=(V,E)$ be a digraph,  $W \subset V$, and $P$ a set of pebbles.
Let $\rho: ( (P \to V) \times E) \to (P \to V)$ be the corresponding transition function. 
Let $G_W=(W,E)$ be the reduced graph. The reduced transition function $\rho_W$ is the transition function on $G_W$.
Namely, $\rho_W: ((P \to W) \times E_W) \to (P \to W)$ is such that, for any valid configuration $\A_W \in P \to W$ and any edge $(u,w)$ with $u \in \A_W(P)$ and $w \notin \A_W(P)$, $\rho_W(\A_W,e)$ is the configuration obtained from $\A_W$ by moving the pebble on vertex $u$ to vertex $v$.
\end{defn}

\begin{rem}
  \label{rem_lift_prop}
  By Proposition~\ref{prop_lift}, for any valid configuration $\A_W \in P \to W$ and $e \in E_W$, there exists a lift $p$ such that  $\rho_W(\A_W, e)=\rho(\A_W,p)$.
\end{rem}

\begin{thm}\label{th:feasibility}
Let $G=(V,E)$ be a digraph and let $<G,\A^s,\A^t, \C>$ be an instance of $\C$-MAPF. 
Let $W \subset V$ be independent on $(G,\C)$ and assume that $\A^s(P),\A^t(P) \subset W$. Let $p=e_1\ldots e_m \in E_W^*$ be a plan and define plan $\hat p=P_1\ldots P_m \in E^*$, where, for $i=1,\ldots,m$, $P_i$ is a lift of $e_i$.
If $p$ solves MAPF $<G,\A^s,\A^t>$, then
$\hat p$ solves $\C$-MAPF $<G,\A^s,\A^t,\C>$.
%   Let $G=(V,E)$ be a strongly connected digraph, $P$ be a set of pebbles on $\F_D$ be a family of subsets defined as in Definition~\ref{family} and $W \in \F_D$ a subset of $V$. Let $\A$ be an initial configuration on $D$ such that $\A(P) \subset W$ and $\tilde{\A}$ be the corresponding initial configuration on $D_W$. Let $a,b\in W$ and $p\in P$ be a pebble on $a$. Then, there is a plan $f_{ab}$ on $D$ consistent with $\C$ and such that $\A[a,b]=\rho(\A,f_{ab})$ if there is a plan $f'_{ab}$ on $D_W$ such that $\tilde{\A}[a,b]=\rho(\tilde{\A},f'_{ab})$.
\end{thm}

\begin{proof}
  Plan $\hat p$ is consistent with $\C$ by Proposition~\ref{prop_lift}.
  By Remark~\ref{rem_lift_prop}, $\rho(\A^s,\hat p)=\A^t$.
\end{proof}

%\begin{proof}
%	Let $f_{ab}'=e_1,...,e_n$ with $e_i \in E_W$ a plan on $D_W$ which moves pebble $p$ from $a$ to $b$. For all $i=1,...,n$, by definition ... ,  $e_i \in E_W$ implies the existence of a path $\pi$ from $a$ to $b$ on $D$ excluding vertices in $W\setminus\{v_1.v_2\}$ and all vertices in $S \in \C$ such that $S \cap (W\setminus \{v_1,v_2\}) \not = \emptyset$. Then, $\pi$ does not contain a node occupied by a pebble or a node associated with an occupied vertex. This prove the existence of a plan $f_{ab}$ on $D$ consistent with $\C$ and such that $\A[a,b]=\rho(\A,f_{ab})$. ANCHE QUI DA SISTEMARE SULLA BASE DELLA NUOVA DEFINIZIONE DI $\C$.
%\end{proof}

\begin{cor}
	\label{cor_finale}
		Let $G=(V,E)$ be a digraph, $\C$ a constraint set, and $\F_G^{\C}$ be the family of subsets defined in Definition~\ref{family}. Let $\A^s$ and $\A^t$ be initial and final configurations, such that $\A^s(P) \cup \A^t(P) \in \F_G^{\C}$. Then, for all $W \in \F_G^{\C}$ such that $\A^s(P)\cup \A^t(P) \subset W$:
	\[<G_W,\A^s,\A^t>  \text{ feasible }\Rightarrow \; <G,\A^s,\A^t, \C> \text{ feasible.}\]
\end{cor}

Corollary~\ref{cor_finale} provides a strategy for finding a solution of $\C$-MAPF. Namely, we look for an independent set of nodes $W$ that contains those associated to the initial and final positions, plus additional ones used for maneuvering, and we solve a standard MAPF problem on the reduced graph. Then, we \textit{lift} the obtained solution to a solution of the original $\C$-MAPF.

As already mentioned, the converse implication ($<G_W,\A^s,\A^t>  \text{ is feasible }\Longleftarrow \; <G,\A^s,\A^t, \C> \text{ is feasible}$) is not true in the general case. For example, if we consider $G=(V,E)$ as in Figure \ref{ex} and $G_{W_1}$ (Figure \ref{ind1}), one can easily verify that, given $\A^s=\{2,4\}$ and $\A^t =\{4,2\}$, $<G,\A^s,\A^t, \C>$ is solvable while $<G_{W_1},s,t> $  is not. Indeed, on $G_{W_1}$ pebbles are forced not to move.\\

In Corollary~\ref{cor_finale}, Assumption $\A^s(P) \cup \A^t(P) \in \F_G^{\C}$ ensures the existence of an independent set containing both source and target vertices. If this assumption is not satisfied, it may still be possible to find a solution of $\C$-MAPF by solving a sequence of two MAPF problems.
Indeed, assume that there exist two sets $W_1 \in \F_G^{\C}$ and $W_2 \in \F_G^{\C}$ such that $\A^s(P) \subset W_1 $, $\A^t(P) \subset W_2 $, and $|W_1 \cap W_2| \geq |P|$, (i.e., their intersection has cardinality greater than the number of pebbles). Let $U$ be a subset of $W_1 \cap W_2$ such that $|U|=|P|$ and let $\A^i:P \to U$ be a valid configuration. Then, we replace the original $\C$-MAPF problem $<G,\A^s,\A^t,\C>$, with a sequence of two $\C$-MAPF problems: $<G_{W_1},\A^s,\A^i>$, which moves pebbles from the source positions to the intermediate positions, and $<G_{W_1},\A^i,\A^t>$ which moves pebbles from the intermediate positions to the targets. The lift of the concatenation of the solutions of these two MAPF problems is a solution of the original $\C$-MAPF problem. This remark leads to the following Corollary of Theorem~\ref{th:feasibility}.

\begin{cor}
	\label{cor_finale2}
	Let $G=(V,E)$ be a digraph and $\F_G^{\C}$ the family of subsets defined in Definition~\ref{family}. Let $\A^s$ and $\A^t$ be initial and final configurations such that $\A^s(P)\in \F_G^{\C}$ and $\A^t(P) \in \F_G^{\C}$.  Then, for all sets $W_1, W_2 \in \F_G^{\C}$ and for all configurations $\A^i$ such that $\A^i(P) \subset W_1 \cap W_2$:
	\begin{gather*}
	<G_{W_1},\A^s,\A^i>  \text{ and }  <G_{W_2},\A^i,\A^t>  \text{ are feasible } \\ \Rightarrow \; <G,\A^s,\A^t, \C> \text{ is feasible.}
	\end{gather*}
\end{cor}

\subsection{Finding an independent set of nodes with largest cardinality}
In view of Corollary~\ref{cor_finale}, we need to find an independent set of nodes $W$ that contains $\A^s \cup \A^t$ (if it does not exist, we can proceed as suggested in Corollary \ref{cor_finale2}). However, it is convenient that $W$ contains the largest number of additional nodes. In fact, the number of holes (i.e., unoccupied vertices) in the reduced MAPF problem $<G_W,\A^s,\A^t>$ is given by $|W|-|\A_s|$, that is the difference between the number of available vertices and the number of pebbles. Various results in literature (see for instance \cite{diBOX,Tass}) show that the feasibility of a MAPF problem depends on the number of holes. Intuitively, the larger is the number of holes, the easier is to find a feasible solution.

This observation leads us to consider the problem of maximizing the cardinality of $W$. %By removing nodes from the original graph $D=(V,E)$, the solution space reduces its dimension and the risk is that  of eliminating too many nodes, i.e., if $|W| <<|V|$,  problem $<D_W,s,t>$ could have no solution. 
We first introduce the notion of a maximal independent set. Intuitively, an independent set is maximal if it is not strictly included into any other independent set.
\begin{defn}
	An independent set $W\in \F_D^{\C}$ is \textbf{maximal} if $ \forall U \in \F_D^{\C}$  such that $W \subset U$, $U = W$.
\end{defn}
Given a digraph $G=(V,E)$ and a constraint set $\C$, we may find various maximal independent sets, of different cardinality. For example, in digraph $D'$ of Figure~\ref{ex}, both $W_1=\{2,4\}$ and $W_2=\{1,3,5\}$ are maximal independent sets. For this reason, it is natural to consider the following problem.
\begin{defn}
  Given a graph $G=(V,E)$, a constraint set $\C$, and a subset $W\subset V$ such that $W \in \C$, the Maximum Independent Set with Additional Constraints ($\C$-MIS) $<G,\C,W>$ consists in finding an independent set that contains $W$, with the largest cardinality.
\end{defn} 
In \cite{paull}, it is proved that finding the largest independent set within a generic ASC is NP-hard. In what follows, we prove that NP-hardness (in fact, strong NP-hardness) holds also for $\C$-MIS (i.e., for the specific
ASC considered in this paper).
We prove the result by a reduction from the Maximum Independent Set (MIS) over graphs, where, given a graph $G=(V,E)$, one searches for the largest independent subset $U$ of $V$, where $U$ is independent (in the classical sense) if $(\forall i,j\in U: i\neq j)\; (i,j)\not\in E$. \\

%As we will see, strong NP-hardness of $\C$-MIS is an almost immediate consequence of strong  NP-hardness of MIS, as stated in the following theorem.

%The maximal independent sets of $G'$ are $S_1 = \{1,5\}$ and $S_2=\{2,3,4\}$: $S_2$ is the largest one and so it is the solution of AC-MIS on $G'$. 

\begin{thm}
	\label{AC_MIS_NP}
	$\C$-MIS is strongly NP-hard.
\end{thm}
\begin{proof}
  The proof is based on a polynomial reduction of the classical Maximum Independent Set (MIS) problem on a graph $G=(V,E)$ to $\C$-MIS.

%Starting from $G$, we define a new (undirected) graph, adding a new node $q$ connected to all nodes of $G$. Formally, the new graph is $G'=(V',E')$, where $V'=V \cup \{q\}$ and $E'=E \cup \{(i,q): \; i \in V\}$. We set a first constraint set $\bar \C$ on $G'$ defined as in (\ref{eq:cq}) where, for each pair $(S,k)\in {\cal Q}$, $k=1$ and $S$ belongs to the collection: 
%$$
%\{\{i,j\}: \; (i,j) \in E\}.
%$$
%We define the overall constraint set as
%\[
%\C=\{U \subset \bar \C: |U|+  |U \cap \{q\}| \leq k\}.
%\]

%Then, we prove the equivalence of the following two problems:

%a) $G$ has a set of $k$ independent vertices (in the classical sense).

%b) there exists a subset of $k$ nodes of $V'$ that is $(G',\C)$ independent.

%Assume that $W \subset V$ is a solution of a). Since $W$ is independent, $W \subset \bar \C$. Moreover, $q \notin W$, so that $W \in \C$. Moreover, each pair $v_1,v_2 \in W$ is connected by path $v_1 q v_2$ (since, $q \notin W$ and $|U \setminus \{v_1,v_2\}|+2=k$). Hence,  $W$ is also a solution of b).

%Conversely, assume that $W \subset V'$ is a solution of b). Note that $q \notin W$. Otherwise, we would have $|W|+  |W \cap \{q\}| = k+1$, which would imply that $W \not\subset \C$. Hence, $W \subset V$ and $W$ is a solution of a) since $W \subset \bar \C$.

%ALTERNATIVA

\noindent Starting from $G$, we define a new (undirected) graph, adding a new node $q$ connected to all nodes of $G$. Formally, the new graph is $G'=(V',E')$, where $V'=V \cup \{q\}$ and $E'=E \cup \{(i,q): \; i \in V\}$. We define a constraint set $\C$ on $G'$ as in (\ref{eq:cq1}) where we introduce:
\begin{itemize}
\item for each $(i,j)\in E$ a triple $(\{i,j\},1,(1,1))\in {\cal Q'}$;
\item the single triple $(V',k,w^{V'})\in {\cal Q'}$, where $w_i^{V'}=1$ for all $i\in V$, while
$w_q^{V'}=2$.
\end{itemize}

\noindent Then, we prove the equivalence of the following two problems:
\begin{itemize}

\item[a)] $G$ has a set of $k$ independent vertices (in the classical sense);

\item[b)] there exists a subset of $k$ nodes of $V'$ that is $(G',\C)$ independent.

\end{itemize}
\noindent Assume that $U \subset V$ is a solution of a). Since $U$ is independent, $|U \cap \{i,j\}|\leq 1$ for all $(i,j)\in E$. Moreover, $q \notin U$, so that $\sum_{i \in U\cap V'} w_i^{V'} = k$. Moreover, each pair $v_1,v_2 \in U$ is connected by path $v_1 q v_2$ (since, $q\notin U$ and $\sum_{i \in [U\setminus \{v_1,v_2\}]\cup \{q\}} w_i^{V'} = k$). Hence,  $U$ is also a solution of b).

Conversely, assume that $U \subset V'$ is a solution of b). Note that $q \notin U$. Otherwise, we would have $\sum_{i\in V'\cap U} w_i^{V'}= k+1$, which would imply that $U \not\in \C$.
Hence, $U \subset V$ and $U$ is a solution of a) since $|U \cap \{i,j\}|\leq 1$ for all $(i,j)\in E$.

%First, note that the solution $S$ of $\C$-MIS cannot contain $q$ because, otherwise, by the definition of $\C$, it could not contain any other node.

%since they are all connected to it: therefore $S \subset V$
%Given $W \subset V$, we need to prove that $W$ is independent (in the classical sense) if and only if $W$ is $(G', \C)$ independent.
%Assume that $W$ is independent, then $W \subset \C$, moreover each Assume that $\C$-MIS problem on $G'$ with constraints $\C$...

%is equivalent to solve the MIS problem on $G$. Indeed, . Moreover, we note that two nodes $i$ and $j$ cannot both belong to the solution of $\C$-MIS only if $(i,j)\in E$. Indeed, if $(i,j)\not \in E$, then they can belong to an independent set since there exists the path $\pi =i \, q \, j$ which connects them.
\end{proof}

\begin{figure}[h!]
	\centering
	\begin{tikzpicture}
		
		\node[shape=circle,draw=green] (A) at (2,1.15) {1};
		\node[shape=circle,draw=black] (B) at (1,0) {2};
		\node[shape=circle,draw=green] (C) at (3,0) {3};
		
		\node[shape=circle,draw=black] (D) at (3,2.3) {4};
		\node[shape=circle,draw=green] (E) at (1,2.3) {5};

		\path (A) edge (B);
		\path (D) edge (C);
		\path (A) edge (D);
		\path (D) edge (E);
		\path (B) edge (E);
		\path (B) edge (C);
		
		%\path [->] (A)  edge[bend right=60]  node[above] {$0$} (D);
		%\path [->] (A)  edge[bend right=80]  node[above] {$0$} (E);

	\end{tikzpicture}
	\caption{Solution of MIS on graph $G$}
	\label{G}
\end{figure}
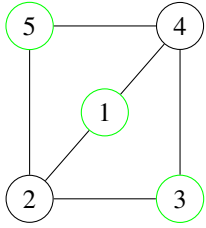

\begin{figure}[h!]
	\centering
	\begin{tikzpicture}
		
		\node[shape=circle,draw=green] (A) at (2,1.15) {1};
		\node[shape=circle,draw=black] (B) at (1,0) {2};
		\node[shape=circle,draw=green] (C) at (3,0) {3};
		
		\node[shape=circle,draw=black] (D) at (3,2.3) {4};
		\node[shape=circle,draw=green] (E) at (1,2.3) {5};
		
		\node[shape=circle,draw=red] (F) at (5,1.15) {6};
		
		\path (A) edge (B);
		\path (D) edge (C);
		\path (A) edge (D);
		\path (D) edge (E);
		\path (B) edge (E);
		\path (B) edge (C);
		
		\path (B)  edge[bend right=60]   (F);
		\path (C)  edge[bend right=20]   (F);
		\path (E)  edge[bend left=60]   (F);
		\path (D)  edge[bend left=20]   (F);
		\path (A) edge (F);
		%\path [->] (A)  edge[bend right=80]  node[above] {$0$} (E);

	\end{tikzpicture}
	\caption{$\C$-MIS on $G'$}
	\label{GG}
\end{figure}

In order to illustrate the result, consider the MIS problem on $G$ (Figure \ref{G}) and the $\C$-MIS problem on $G'$ (Figure \ref{GG}) with $\mathcal{S}= \{\{i,j\}\ :\  (i,j)\in E'\}$.  
It is easily observed that $S=\{1,3,5\}$ is the largest solution of MIS on $G$. We note that one of the maximum solutions of $\C$-MAPF on $G$ would be $T=\{1,3\}\subset S$. If we consider graph $G'$ of Figure \ref{GG}, node 5 can also be added to the independent set $T$, as it is connected to both 1 and 3 by a free path through 6. Therefore, the solutions of the two problems on the two different graphs are equivalent.

\subsection{Heuristics for the $\C$-MIS problem}
\label{heuristic}
Paull and Unger's procedure, presented in~\cite{paull}, finds the maximal independent sets, and, in particular, the largest one, within a general ASC. As already mentioned, reference~\cite{paull} also proves that detecting the largest maximal independent set is NP-hard. Hence, we can apply Paull and Unger's procedure to solve $\C$-MIS, but only for very small instances.

In Theorem~\ref{AC_MIS_NP}, we proved that also $\C$-MIS is (strongly) NP-hard.
In the following, we introduce polynomial-time heuristic algorithms for obtaining a good quality, sub-optimal solution of $\C$-MIS.
We describe an iterative procedure that finds a maximal independent set $M$ over a graph $G=(V,E)$. In the procedure, described in Algorithm \ref{MIS}, we first set $M=W$, where $W$ is some initial set belonging to $\C$. 
Set $W$ can be equal to $\A^s\cup A^t$ in view of Corollary~\ref{cor_finale} or to $\A^s\cup A^i$ (or $\A^i\cup A^t$) in view of Corollary~\ref{cor_finale2}.
Then, at each iteration, it searches for a vertex $v$ such that $M \cup \{v\}$ is independent, among all the nodes  of set $\Delta:=V \setminus \mathcal{E}$. Set $\mathcal{E}$ contains the nodes that can no longer be added and, in particular, $M \cup \mathcal{L}(M) \subset \mathcal{E}$ where

\[   \mathcal{L}(M) =  \{w \in V : \;  \; M \cup\{w\} \not \in \C\}.\]

 If such vertex does not exist, $M$ is a maximal independent set and the procedure terminates. Otherwise, $v$ is added to set $M$, $\mathcal{L}(M)$ is updated,  and a new iteration
is performed. 
To avoid repetition in the following iterations, all selected nodes in this iteration are added to $\mathcal{E}$.

\begin{obser}
	\label{compl}
	Let $G=(V,E)$ be a directed graph, $M\subset V$ a subset of nodes, and $\C$ defined as in (\ref{eq:cq1}) the constraint set. Then, building the reduced graph $G_M =(M,E_M)$ has worst time complexity $O(n^2 \cdot (n^2+h))$, where $n=|V|$ and $h=|\mathcal{Q}|$.
\end{obser}
\begin{proof}
 The procedure involves two steps for each couple of nodes $(u,v)\in V \times V$:
\begin{itemize}
\item constructing $G_M^{u,v}=(V_{M}^{u,v},E_{M}^{u,v})$;
\item checking whether there exists a path from $u$ to $v$ on $G_M^{u,v}$.
\end{itemize}

\noindent The latter step can be done by  depth-first search, which takes $O(|V_{M}^{u,v}| + |E_{M}^{u,v}|)$. However, in the worst case of a complete graph, it has time complexity $O(n^2)$. As for the former, if $\C$ is defined as in (\ref{eq:cq1}), it takes $O(h)$, where $h=|\mathcal{Q}|$. Indeed, to build this graph, for each $(S_i,k_i,w_i^S)\in \mathcal{Q}$ we have to check which nodes $w\in V$ are such that $\sum_{j\in S_i \cap M \setminus \{u,v\}} w_j^S \leq k_i$.

 To do that, we propose the following procedure. For each node $x\in V$ we define a vector $\bar{x}$ of length $h$ where, for each $i=1,...,h$, 

\begin{equation}
	\bar{x}_i = \Bigg\{ 
		\begin{array}{ll}
		w_{x}^S & \text{ if } x \in S_i, \\
		0 & \text{otherwise}.
		\end{array} 
\end{equation}
Moreover, we define a vector $m$ of length $h$, which represents at each iteration which is the occupied capacity of the nodes that stay in the intersection between $M$ and $S_i$:

\[	m_i = \sum_{j\in S_i \cap M}  w_j^S.\]

This vector is initialized with the null vector, and it is updated iteratively each time a vertex $x$ is added to $M$:

\[m_i = m_i + \bar{x}_i, \quad \forall i=1,...,h.\]

Therefore, checking  whether $\sum_{j\in S_i \cap M \setminus \{u,v\}} w_j^S \leq k_i$ is equivalent to check whether  
$(m_i-\bar{u}_i-\bar{v}_i+ \bar{x}_i) \leq k_i$, which takes $O(h)$.

\end{proof}

\begin{prop}
	If $\C$ is defined as in  (\ref{eq:cq1}), Algorithm \ref{MIS} has worst time complexity $O(n^3 \cdot (n^2+h))$, where $n=|V|$ and h = $|\mathcal{Q}|$.
\end{prop}

\begin{proof}
By Observation \ref{compl}, at each iteration of Algorithm \ref{MIS}, building the reduced graph $G_M=(M,E_M)$ has worst time complexity $O(n^2 \cdot (n^2+h))$. Moreover,  checking whether $G_{M}$ is strongly connected is dominated by its construction and takes $O(|M| + |E_M|)$. However, in the worst case of a complete graph, it has time complexity $O(n^2)$. Therefore, checking whether $M$ belongs to $\F_{G}^{\C}$ takes $O(n^2 \cdot (n^2+h))$. Since each node is selected at most once, then the operation of checking whether a set is independent is perfomed at most $n$ times. It follows that Algorithm \ref{MIS} has worst time complexity $O(n^3 \cdot (n^2+h))$. 
\end{proof}

\RestyleAlgo{ruled}
\begin{algorithm}
	\caption{Maximal independent set}
	\label{MIS}
	Let $M$ be a (possibly empty) independent set\;%  = \emptyset$ (or $M=s \cup t$) \;
   
    $MaxSetFound = false$\;
    
    $\mathcal{E} = M \cup \mathcal{L}(M)$;
    
     $\Delta = V\setminus \mathcal{E}$\;
     
    \While{$MaxSetFound = false$}{

    	 	$NewNodeFound = false$;
    	 	
    \While{$\Delta \not = \emptyset$ and $NewNodeFound = false$}{
    Select $v \in \Delta$;
    
    $N = M \cup \{v\}$;
    
    \If{$N\in \F_G^{\C}$}{
    
    	$M = N$\;
    	$NewNodeFound = true$;
    	
    	Update $\mathcal{L}(M)$;
    	
    	$\mathcal{E} = \mathcal{E} \cup \mathcal{L}(M)$;
    }
 $\mathcal{E} = \mathcal{E} \cup \{v\}$;

     $\Delta = \Delta \setminus \mathcal{E}$;	
    }
\If{$NewNodeFound = false$}{
	
$MaxSetFound = true;$}
}
\end{algorithm}
Different heuristics can be obtained by changing the rule to select node $v$.
The first one is the \textit{Random} approach, which, at each step, selects the next node randomly. The second one is the \textit{Greedy} approach, where
node $v$ is selected through a selection criterion $\psi: V \times \C  \rightarrow \mathbb{R}$:
\[  v = \arg \max_{w \in V \setminus M} \psi(w,M).\]

The selection criterion can be defined in different ways. If the constraint set $\C$ is defined  as in (\ref{eq:cq}), an idea is to give precedence to nodes that are less constrained by other nodes. In this case, we can define the selection criterion as follows

\begin{equation}
	\label{psi2}
	   \psi(v,M) := | \{w\in V :\; M \cup\{v,w\}\in \C \}|,	     
\end{equation}
in order to choose the node that would allow to further expand set $M$ in the following steps.

The selection criterion may also take into account the form of $\C$. For example,  let
$\C$ be defined as in (\ref{eq:cq}) where, for each pair $(S,k)\in {\cal Q}$, $k=1$ and $S$ belongs to the collection: 
\begin{equation}
	\label{s}
	\mathcal{S}=\{\{i,j\}: \; (i,j) \in E\}.
\end{equation} 
Then, if we indicate with $G \setminus \mathcal{E}$ the subgraph of $G$ obtained by erasing the nodes in $\mathcal{E}$, a possible selection criterion is the inverse of the degree on $G \setminus \mathcal{E}$:
\begin{equation}
	\label{deg}
	\psi(v,M) := \frac{1}{\deg_{G \setminus \mathcal{E}}(v)}.
\end{equation}

Indeed, nodes with a lower degree are nodes bounded to a smaller number of other nodes. Therefore, adding a vertex with a lower degree to the independent set usually means having more choices in the following steps.

%Another approach is the Greedy Randomized Adaptive Search Procedure (GRASP, see \cite{FeoResende89}), a variant of the greedy approach where a probability $\pi$, associated to the criterion $\psi$, is associated to each node. The value $\pi(v)$ represents the probability for node $v$ to be selected. We impose that $\pi$ is monotone with respect to $\psi$, in particilar the lower $\psi$ is, the larger the probability of selecting the node is:

%\[  \psi(v) \leq \psi(w) \iff \pi(v) \geq \pi(w).\]

%If $\psi$ is defined as in (\ref{deg}), we consider $Deg = (\text{Deg}_1, \,...\,,\text{Deg}_k)$ the ascending ordered sequence of all the degrees of the nodes of graph $G$ (i.e., $\text{Deg}_i<\text{Deg}_{i+1}$ for all $i=1,...,k-1$), and we choose a descending ordered sequence of probabilities $p=(p_1,\,...\,,p_k)$ (i.e., $1>p_i>p_{i+1}>0$ for all $i=1,...,k-1$) such that $\sum_{i=1}^{k}p_i=1$. We define $\pi$ in this way

%\begin{equation}
	%\label{GRASP}
	%\text{if} \; \; \psi(v) = \text{Deg}_i\;\; \Longrightarrow \;\; \pi(v) = \frac{p_i}{n_i}, \quad \text{ where } n_i=\symbol{35}\{w \in V : \, deg(w)=Deg_i\}.	
%\end{equation}

\section{Experimental results}
\subsection{Real-life applications examples of $\C$-MAPF}
\begin{figure}[h!]
	\centering
	\includegraphics[width=0.7\columnwidth,trim={2.5cm 1.8cm 2cm 1.5cm},clip]{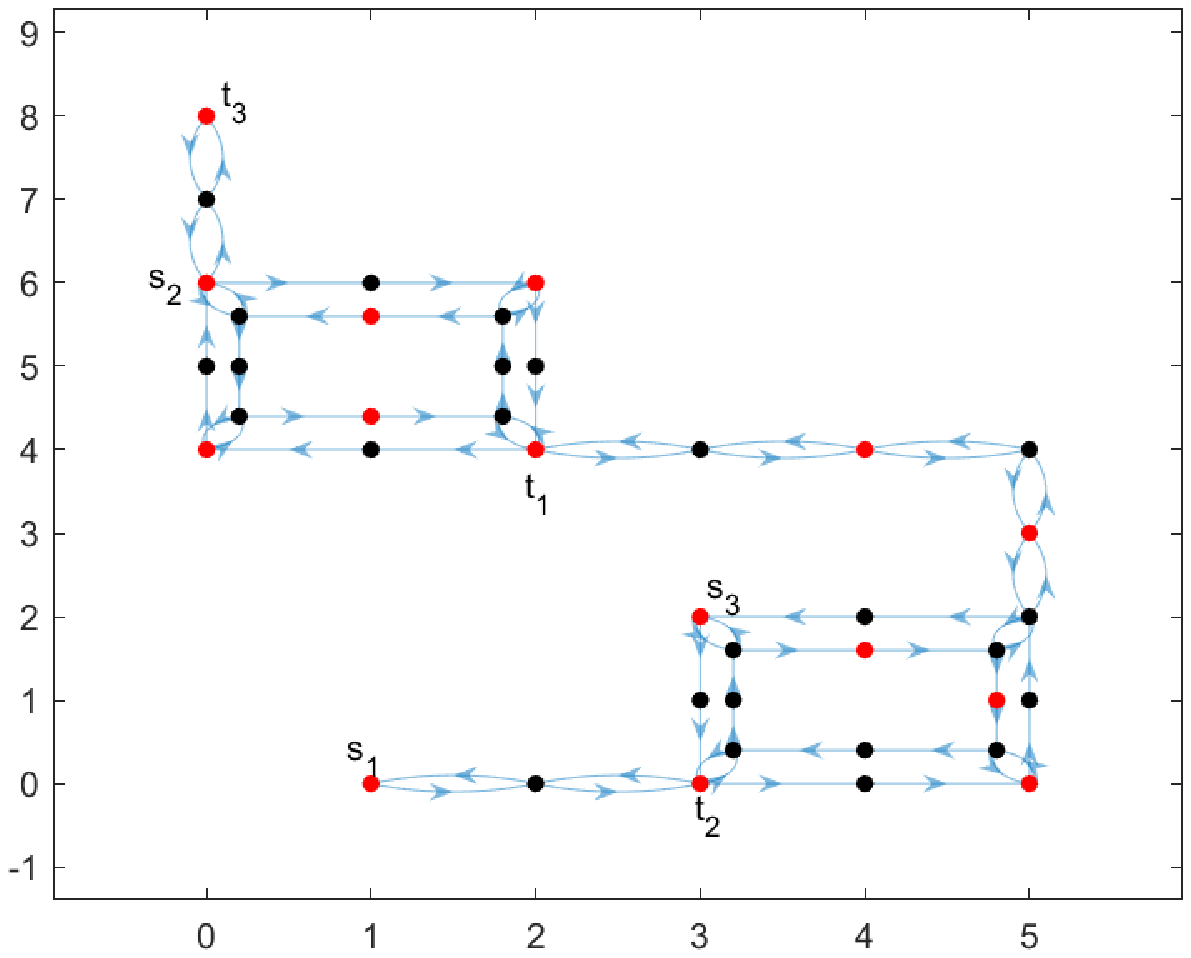}
	\includegraphics[width=0.7\columnwidth,trim={2.5cm 1.8cm 2cm 1.5cm},clip]{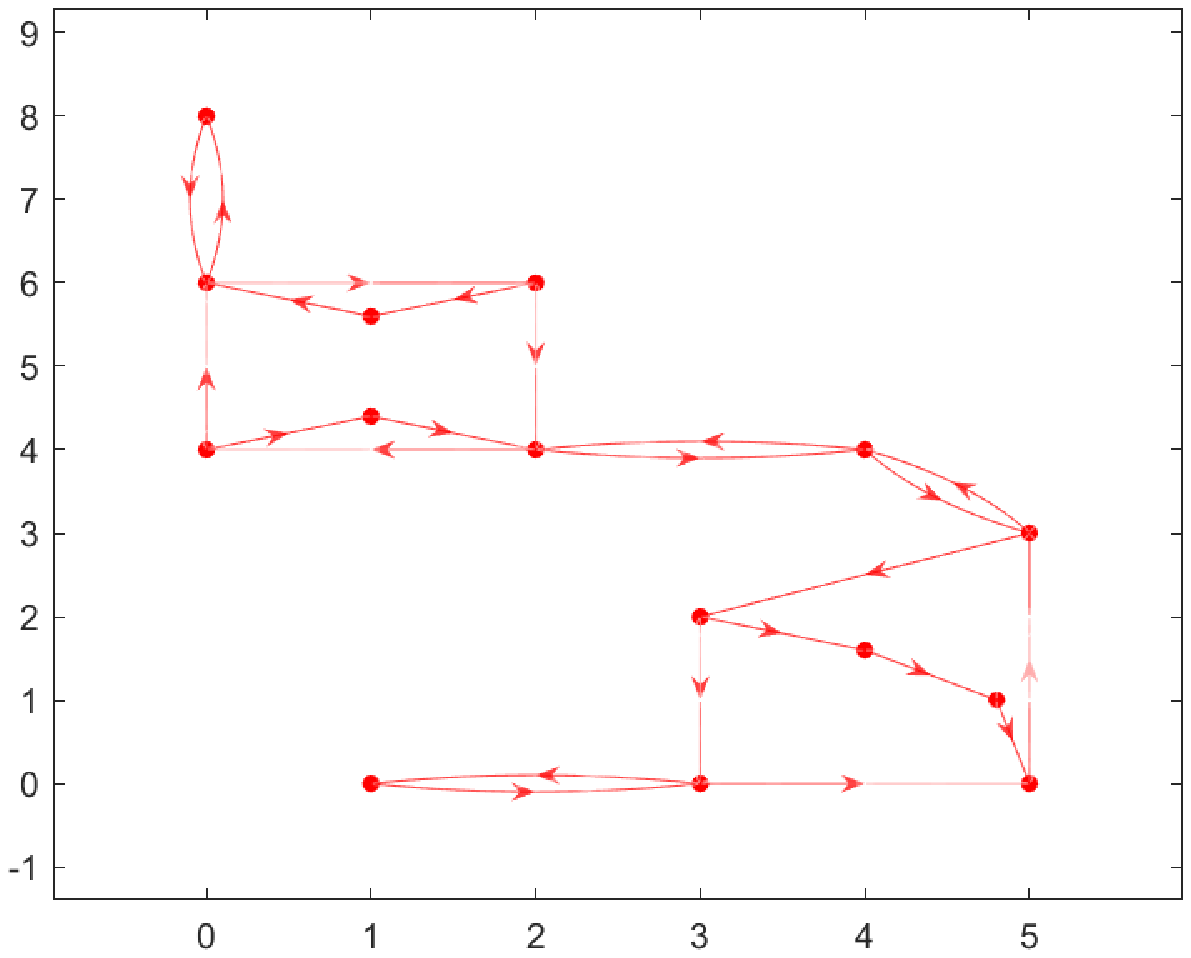}	
	\caption{Example of graph representing an industrial scenario and the corresponding reduced graph.}
	\label{fig:mag}
\end{figure}

Consider the digraph $G=(V,E)$ in Figure~\ref{fig:mag}. It represents a small warehouse with three agents. Agent 1 has to move from $s_1$ to $t_1$, agent 2 from $s_2$ to $t_2$, and agent 3 from $s_3$ to $t_3$. To maintain a safety distance, we require that agents cannot occupy adjacent vertices at the same time. This is equivalent to consider a constraint set $\C$ defined as in (\ref{eq:cq}), where, for each pair $(S,k) \in Q$, $k=1$ and $S \in \mathcal{S}$ (where $\mathcal{S}$ is defined as in (\ref{s})). Note that the union of the sources $S = \{s_1, \,s_2,\, s_3\}$ and of the targets $T = \{t_1, \,t_2,\, t_3\}$ is an independent set on $(G,\C)$. To reduce $\C$-MAPF to MAPF, we use Algorithm 1, setting $W=S \cup T$, to find a maximal independent set. In this way, we are able to obtain a reduced graph $G'$. We can solve the (unconstrained) MAPF problem on $G'$. Here, we do not discuss the solution of MAPF, since it is extensively discussed in literature. Finally, we lift the obtained MAPF solution to a solution of the original $\C$-MAPF.
%we can assume that AGV1, AGV2 and AGV3 occupy single nodes, and we can use any algorithm for MAPF problem to
%find a path for each agent that allows them to reach their destination avoiding any collision.
In Figure~\ref{fig:mag}, we show the best maximal independent set on $G$, containing $W$, obtained after 100 iterations of Algorithm~1 with the \textit{Random} approach, and the correponding reduced graph $G'$.

As a second example, we considered a real-life warehouse layout, provided by packaging
company Ocme S.r.l., based in Parma, Italy. The set of constraints $\C$ takes into account the dimensions of the AGV vehicles. The set $\C$ is defined as in  (\ref{eq:cq}), where for each pair $(S,k) \in Q$, $k=1$ and $S \in \mathcal{S}$, with $\mathcal{S}$ defined as follows. For each node $v \in V$ on the graph, the company provides a set of edges $\tilde{E}_v$ that are bounded by that node. We define $\mathcal{S}$ as the set of triplets of vertices $(v,i,j)$ such that $(i,j)\in \tilde{E}_v$:

\[  \mathcal{S}=\{(v,i,j) : \;v,i,j\in V,\;(i,j)\in \tilde{E}_v\}.\]
%\begin{itemize}
%	\item $k=1$;
%	\item $S$ is a set of two vertices $i$, $j$ such that if one of them is occupied by a vehicle then the other one, if occupied by another vehicle, would cause a collision or an interference between the vehicles themselves. (NON E' ESATTAMENTE COSI' ANDREBBE CHIARITO)
%\end{itemize} 

\begin{figure}[h!]
	\centering
	\includegraphics[width=0.8\columnwidth,trim={5cm 2cm 3cm 2cm},clip]{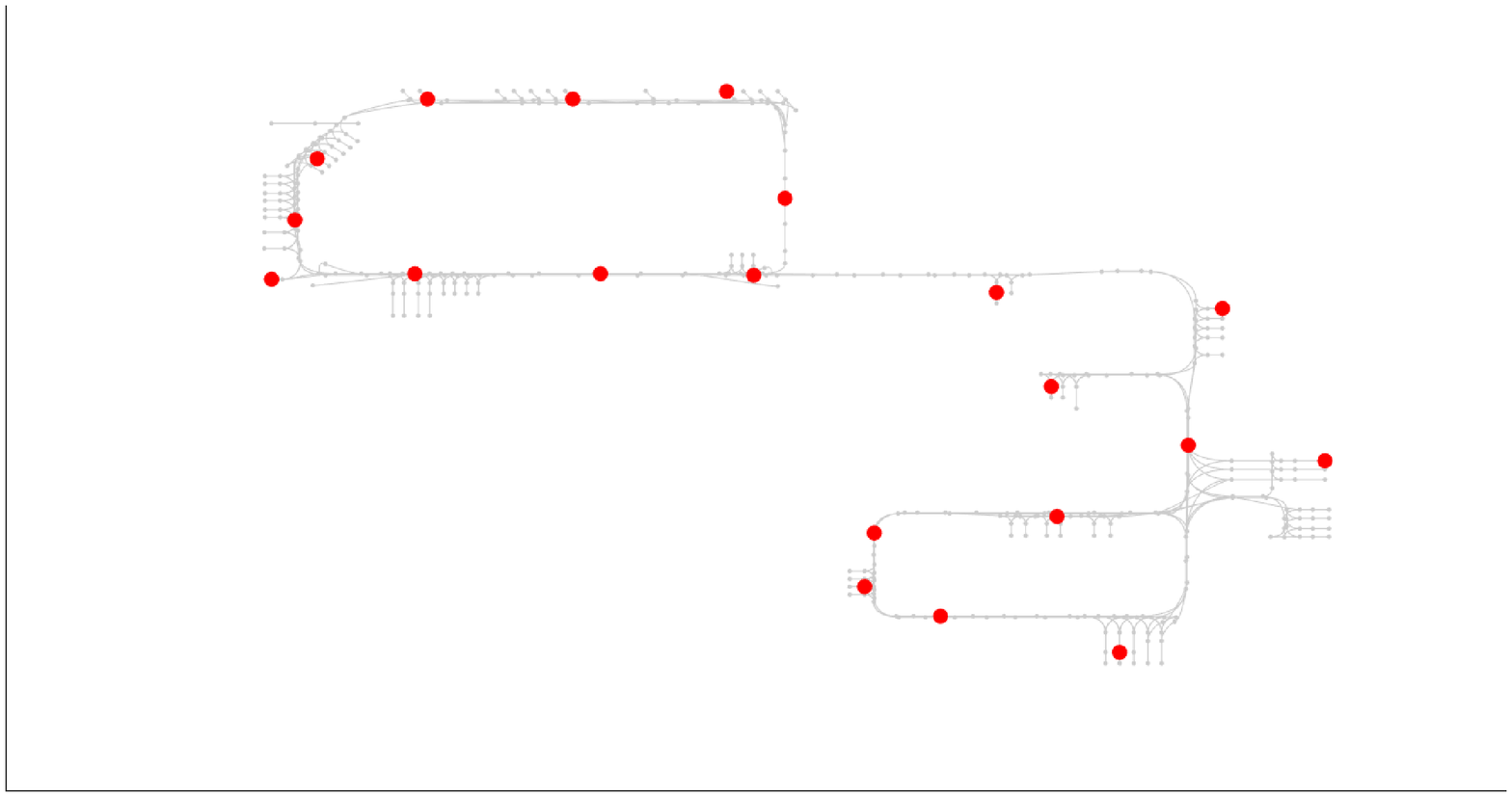}
		\includegraphics[width=0.7\columnwidth,trim={5cm 2cm 3cm 2cm},clip]{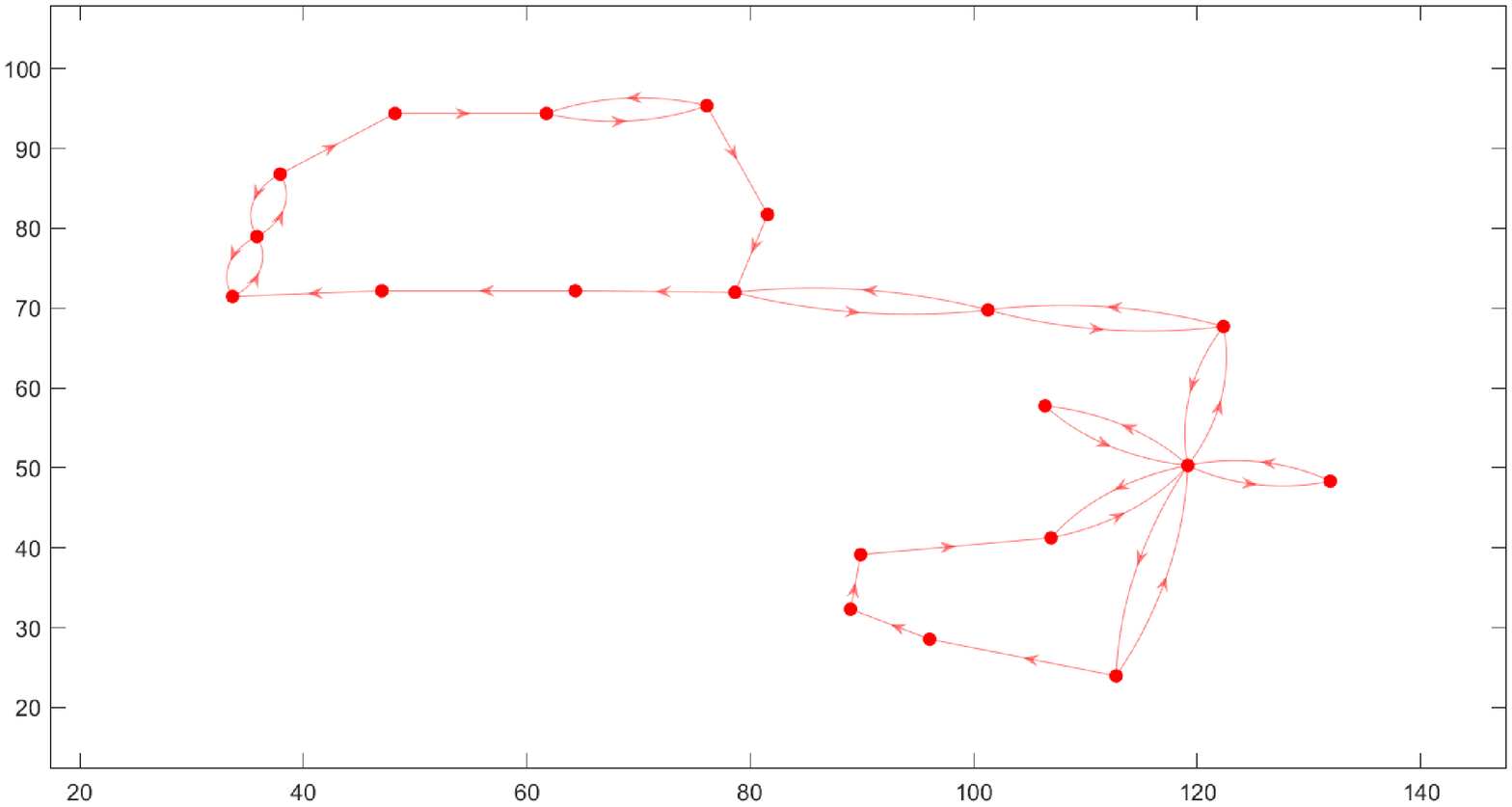}	
	\caption{ A real-life warehouse layout, and the corresponding reduced graph.}
	\label{fig:pedrollo}
\end{figure}

We ran 100 times the random variant of Algorithm~\ref{MIS}, and we computed a maximal independent set of cardinality equal to 20. The red vertices of Figure~\ref{fig:pedrollo} show this independent set. %Note that these positions are well distributed in the warehouse, and therefore respect the safety distances between the AGVs, taking into account their
%dimension. 

%\begin{figure}[h!]
	%\centering
	%\includegraphics[width=0.7\columnwidth,trim={5cm 2cm 3cm 2cm},clip]{pedrollo_ridotto.eps}
	%\caption{Pedrollo.}
	%\label{fig:pedrollo}
%\end{figure}%
These results suggests that on a real-life industrial scenario, the number of independent vertices is sufficiently large to allow solving $\C$-MAPF problems with a significant number of agents. In particular, taking into account the number of nodes of the two graphs and the length of the  corridors, and using the results of \cite{diSC,Tass,PMT}, it is possible to calculate that in the reduced graphs of Figure \ref{fig:mag} we can solve all MAPF instances with up to 9 agents, while in the reduced graph of Figure \ref{fig:pedrollo}, all MAPF instances with up to 14 agents.

\subsection{$\C$-MIS problem on square grid graphs}
\begin{figure}[h!]
	\centering
	\begin{subfigure}[b]{0.45\textwidth}
		\centering
	\begin{tikzpicture}
		
		\node[shape=circle,draw=blue,line width=1mm] (A) at (-1,0) {4};
		\node[shape=circle,draw=black] (B) at (1,0) {5};
		\node[shape=circle,draw=blue,line width=1mm] (C) at (3,0) {6};
		
		\node[shape=circle,draw=yellow,line width=1mm] (D) at (-1,2) {1};
		\node[shape=circle,draw=yellow,line width=1mm] (F) at (3,2) {3};
		\node[shape=circle,draw=blue,line width=1mm] (E) at (1,2) {2};
		
		\node[shape=circle,draw=yellow,line width=1mm] (G) at (-1,-2) {7};
		\node[shape=circle,draw=yellow,line width=1mm] (I) at (3,-2) {9};
		\node[shape=circle,draw=blue,line width=1mm] (H) at (1,-2) {8};
		
		\path (A) edge (B);
		%\path (D) edge (C);
		\path (A) edge (D);
		\path (D) edge (E);
		\path (B) edge (E);
		\path (B) edge (C);
		\path (F) edge (C);
		\path (I) edge (C);
		\path (F) edge (E);
		\path (G) edge (H);
		\path (I) edge (H);
		\path (F) edge (C);
		\path (B) edge (H);
		\path (F) edge (C);
		\path (G) edge (A);
		%\path [->] (A)  edge[bend right=60]  node[above] {$0$} (D);
		%\path [->] (A)  edge[bend right=80]  node[above] {$0$} (E);

	\end{tikzpicture}
	\caption{Grid Graph of $3\times 3$. $W_{1}=\{1,3,7,9\}$ and $W_2=\{2,4,6,8\}$ are the two optimal solutions of $\C$-MIS.}
	\label{G3}
\end{subfigure}
\begin{subfigure}[b]{0.43\textwidth}
	\centering
	\begin{tikzpicture}[scale =0.7]
		
		\node[shape=circle,draw=black] (A) at (-1,0) {6};
		\node[shape=circle,draw=blue,line width=1mm] (B) at (1,0) {7};
		\node[shape=circle,draw=black] (C) at (3,0) {8};
		\node[shape=circle,draw=black] (A1) at (5,0) {9};
		\node[shape=circle,draw=black] (B1) at (7,0) {10};
		
		\path (C) edge (A1);
		\path (A1) edge (B1);
		
		\node[shape=circle,draw=blue,line width=1mm] (D) at (-1,2) {1};
		\node[shape=circle,draw=blue,line width=1mm] (F) at (3,2) {3};
		\node[shape=circle,draw=black] (E) at (1,2) {2};
		\node[shape=circle,draw=black] (D1) at (5,2) {4};
		\node[shape=circle,draw=blue,line width=1mm] (F1) at (7,2) {5};
		
		\path (F) edge (D1);
		\path (D1) edge (F1);
		\path (C) edge (A1);
		\path (A1) edge (D1);	
		\path (F1) edge (B1);

		\node[shape=circle,draw=black] (G) at (-1,-2) {11};
		\node[shape=circle,draw=black] (I) at (3,-2) {13};
		\node[shape=circle,draw=black] (H) at (1,-2) {12};
		\node[shape=circle,draw=blue,line width=1mm] (G1) at (5,-2) {14};
		\node[shape=circle,draw=black] (H1) at (7,-2) {15};
		
		\path (I) edge (G1);
		\path (G1) edge (H1);		
		\path (G1) edge (A1);	
		\path (B1) edge (H1);

		\node[shape=circle,draw=blue,line width=1mm] (J) at (-1,-4) {16};
		\node[shape=circle,draw=blue,line width=1mm] (L) at (3,-4) {18};
		\node[shape=circle,draw=black] (K) at (1,-4) {17};
		\node[shape=circle,draw=black] (M) at (5,-4) {19};
		\node[shape=circle,draw=blue,line width=1mm] (N) at (7,-4) {20};
		
		\path (J) edge (K);
		\path (K) edge (L);		
		\path (L) edge (M);	
		\path (M) edge (N);
		\path (J) edge (G);
		\path (K) edge (H);		
		\path (L) edge (I);	
		\path (M) edge (G1);
		\path (N) edge (H1);

		\node[shape=circle,draw=black] (O) at (-1,-6) {21};
		\node[shape=circle,draw=black] (Q) at (3,-6) {23};
		\node[shape=circle,draw=blue,line width=1mm] (P) at (1,-6) {22};
		\node[shape=circle,draw=blue,line width=1mm] (R) at (5,-6) {24};
		\node[shape=circle,draw=black] (S) at (7,-6) {25};
		
		\path (J) edge (O);
		\path (K) edge (P);		
		\path (L) edge (Q);	
		\path (M) edge (R);
		\path (N) edge (S);
		\path (O) edge (P);		
		\path (P) edge (Q);	
		\path (Q) edge (R);
		\path (R) edge (S);

		\path (A) edge (B);
		%\path (D) edge (C);
		\path (A) edge (D);
		\path (D) edge (E);
		\path (B) edge (E);
		\path (B) edge (C);
		\path (F) edge (C);
		\path (I) edge (C);
		\path (F) edge (E);
		\path (G) edge (H);
		\path (I) edge (H);
		\path (F) edge (C);
		\path (B) edge (H);
		\path (F) edge (C);
		\path (G) edge (A);
		%\path [->] (A)  edge[bend right=60]  node[above] {$0$} (D);
		%\path [->] (A)  edge[bend right=80]  node[above] {$0$} (E);

	\end{tikzpicture}
	\caption{Grid Graph of $5\times 5$. 
		 $W_{1}=\{1,3,5,7,14,16,18,20,$ $22,24\}$ is an optimal solutions of $\C$-MIS.}
	\label{G5}
\end{subfigure}
\caption{$\C$-MIS on grid graphs.}

\end{figure}
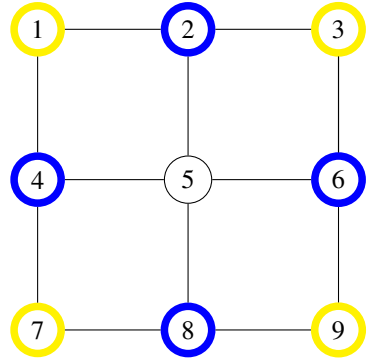
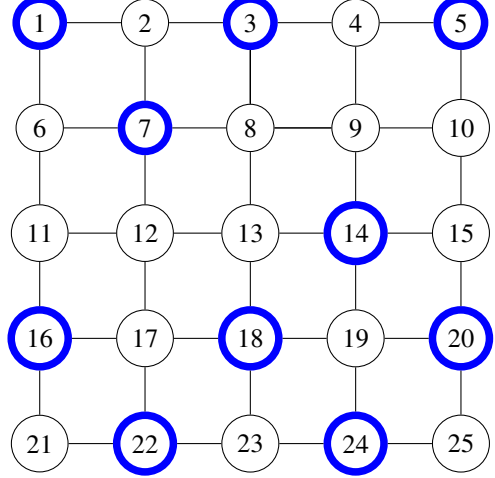

We focus on the problem of finding the independent set of maximum cardinality. In particular, we consider square grid graphs, as in Figures~\ref{G3},~\ref{G5}. The constraint set $\C$ is defined as in (\ref{eq:cq}), where, for each pair $(S,k) \in Q$, $k=1$ and $S \in \mathcal{S}$ (where $\mathcal{S}$ is defined as in (\ref{s})). In other words, at any time, at most one agent can be positioned in the two vertices associated to each edge.

First, we found the independent sets of maximal cardinality by Paull and Unger's algorithm.
Due to the computational complexity of this algorithm (recall that $\C$-MIS is NP-hard), we could not solve instances on grid larger than $5 \times 5$ vertices in a reasonable time. Indeed, solving the $5 \times 5$ required more than 20 hours. 

Figure~\ref{G3} shows the two optimal solutions on the $3 \times 3$ grid, while Figure~\ref{G5} shows one optimal solution $W_1$ for the $5 \times 5$ grid. 
The optimal solution of the latter is unique up to the isometries belonging to the dihedral group $D_4$. Indeed, all the other seven optimal solutions can be found with the composition of  $k\pi/4$ rotations and symmetries with respect to the axes.

We also tested the heuristic algorithms presented in Section~\ref{heuristic} to find independent vertices (without guarantee of optimality) on square grid graphs with $n \times n$ vertices, for $n =2,\ldots,12$.
In particular, we implemented Algorithm~\ref{MIS} both with the purely random selection rule and with the greedy rule (defined as in (\ref{deg})).
 
\begin{figure}[h!]
	\centering
	\includegraphics[width=1\columnwidth]{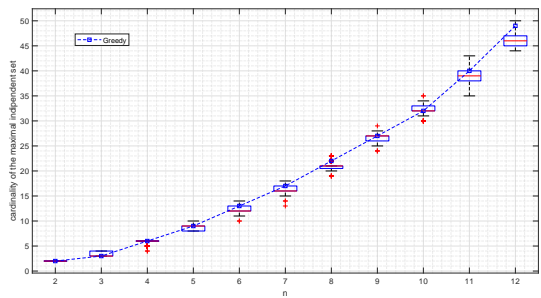}
	\caption{Comparison of random and greedy.}
	\label{fig:RG}
\end{figure}

 For each graph, we 
ran the random algorithm 100 times, while we ran the greedy version only once, since the greedy rule is deterministic.
Figures~\ref{fig:RG} presents the distribution of the cardinality of the independent sets returned by the random approach, compared with that of the greedy algorithm (the blue square is the cardinality of the greedy solution).
Note that the single solution returned by the greedy approach is usually better than the average solution returned by the random one.  However, the best solution over the 100 runs of the first approach is usually better than the single solution returned by the greedy approach, as shown in Figures~\ref{table}, in which we compare these values and that one of the optimal solution returned by Paull and Unger's algorithm.

These experimental results show that the heuristc algorithms find suboptimal solutions of good quality, 
that do not differ much from each other and from the best solutions.
We are not able to certify that we have found the maximum cardinality independent set for grid graphs of large dimensions, but these algorithms appear to find a good approximation that allows not to reduce too much the instances of the problem.

\begin{figure}[h!]
\centering
\begin{tabular}{|r|r|r|r|}
	
	\hline
	n & Random & Greedy &  optimal solution \\
	\hline
	2 & 2 & 2 &  2 \\
	3 & 4 & 3 &  4 \\
	4 & 6 & 6 &  6 \\
	5 & 10 & 9 &  10 \\
	6 & 14 & 13 &  - \\
	7 & 18 & 17 &  - \\
	8 & 23 & 22 &  - \\
	9 & 29 & 27 &  - \\
	10 & 35 & 32 &  - \\
	11 & 43 & 40 & - \\
	12 & 50 & 49 &  - \\
	\hline

\end{tabular}
\caption{Comparison of maximum cardinalities of independent sets found with Random, Greedy, and the optimal algorithm.}	
\label{table}
\end{figure}

\section{Conclusion and future works}
 %{\bf Conclusion and future works}

We introduced $\C$-MP and $\C$-MAPF, generalizations of classic MP and MAPF, which take into account additional constraints on the nodes of the graph. We proved that the problem of finding a feasible solution is NP-hard, and we proposed a reduction method from these new problems to the classic ones. However, this method does not allow solving some $\C$-MP and $\C$-MAPF instance, since the reduced graph excludes some solutions of the original problem. 
This suggests an idea for a possible future development, in which  we could study another reduction method that guarantees to find a feasible solution for a larger class of $\C$-MP and $\C$-MAPF instances, preserving polynomiality.

Moreover, we have dealt with the problem of how to construct the reduced graph and we have proposed some heuristics to find a solution of the $\C$-MIS problem in polynomial time.
 In the future, we could study more efficient heuristics for the $\C$-MIS problem.
%As a third experiment, we generated random graphs with a number of nodes that ranges from 4 to 140. For each number of nodes, we generated a set of 50 graphs. In order to generate test graphs with multiple biconnected components, we used a procedure presented in \cite{diSC}. On each graph we generate randomly a constraint set $\C$ defined as in ....In this simulation we compare the Greedy algorithm with the selection criterion defined as in (\ref{psi2}) and the random approach (Algorithm 1). For the latter, we perform 20 iteration for each graph in order to find the best maximal independent set.  The results shown in Figure \ref{fig:mv} and \ref{fig:mv2} confirm that $\psi$ defined as in (\ref{psi2}) is an efficient selection criterion, since it prefers nodes with fewer constraints and this allows to generate a larger independent set.

\bibliographystyle{siamplain}

\begin{thebibliography}{9}
	\bibitem{diSC} S. Ardizzoni, I. Saccani, L. Consolini,  M. Locatelli, Multi-Agent Path Finding on Strongly Connected Digraphs,{ \em arXiv}, \url{https://arxiv.org/abs/2209.04286}, (2022)
\bibitem{robust} D.
Atzmon, R. Stern, A. Felner, G. Wagner, R. Bartak,
and N.-F. Zhou, Robust multi-agent path finding, {\em
International Conference on Autonomous Agents and Multi
Agent Systems (AAMAS)}, 1862–1864, (2018)
\bibitem{Balas75} E. Balas, Facets of the knapsack polytope, {\em Mathematical Programming}, 8, 146--164 (1975)
\bibitem{diBOX} A. Botea, P. Surynek, Multi-agent path finding on strongly biconnected digraphs, {\em Journal of Artificial Intelligence Research} 62, 273–314 (2018)
\bibitem{FeoResende89} T.A. Feo and M. G. C. Resende, A probabilistic heuristic for a computationally difficult set covering problem, {\em Operations Research Letters}, 8 (2), 67--71 (1989)
\bibitem{H} W. Honig, T. K. S. Kumar, L. Cohen, H. Ma, H. Xu, N. Ayanian,
	 and S. Koenig,  Multi-agent path finding with kinematic
	constraints, {\em Proceedings of the Twenty-Sixth International Conference on
Automated Planning and Scheduling (ICAPS 2016)}, 477–485
(2016)
\bibitem{Tass} M. Khorshid , R.C. Holte, N.R. Sturtevan, A polynomial-time algorithm for non-optimal multi-agent path finding, \textit{The Fourth Annual Symposium on Combinatorial Search (SoCS'11)}, 76-83 (2011)
\bibitem{PMT} A. Krontiris, R. Luna, K.E. Bekris, From Feasibility Tests to Path Planners for Multi-Agent Pathfinding, {\em Sixth Annual Symposium on Combinatorial Search (SoCS)}, Vol. 4 No. 1, 114-122 (2013)
\bibitem{paull} E. L. Lawler, J. K. Lenstra and A. H. G. Rinnooy Kan, Generating all maximal independent sets: NP-hardness and polynomial-time algorithms, {\em Society for Industrial and Applied Mathematics}, (1980)
\bibitem{la} J. Li, P. Surynek, A. Felner, H. Ma T.K. Satish Kumar Sven Koening, Multi-Agent Path Finding for Large Agents, 
{\em Proceedings of the AAAI Conference on Artificial Intelligence}, 33(01), 7627-7634 (2019)
\bibitem{GMP1R}
C. H. Papadimitriou, P. Raghavan, M. Sudan, and
H. Tamaki,  Motion planning on a graph, {\em Foundations of
Computer}, 511-520, (1994)
 \bibitem{cbs}
G. Sharon, R. Stern, A. Felner,  N. R. Sturtevant, Conflict-based search for optimal multi-agent pathfinding, {\em Artificial Intelligence}, 219, 40-66 (2015)
\bibitem{hca}
D. Silver, Cooperative pathfinding, {\em AIIDE'05: Proceedings of the First AAAI Conference on Artificial Intelligence and Interactive Digital Entertainment}, 117-122 (2005)
\bibitem{MAPF} R. Stern et al., Multi-Agent Pathfinding: Definitions, Variants, and Benchmarks, {\em Proceedings of the Twelfth International
Symposium on Combinatorial Search (SoCS 2019)}, 151-158 (2019)
\bibitem{feasibility}
Z. Wu, S. Grumbach, Feasibility of motion planning on acyclic and strongly connected
directed graphs, {\em Discrete Applied Mathematics}, 158(9), 1017-1028 (2010)
\bibitem{WuGru09} Z. Wu, S. Grumbach, Feasibility of Motion Planning on Directed Graphs, {\em In TAMC 2009. Lecture Notes in Computer Science}, vol 5532. Springer.
\bibitem{np} J. Yu, S. M. LaValle, Structure and Intractability of optimal multi-robot path planning on graphs, 
{\em Proceedings of the AAAI Conference on Artificial Intelligence}, 27(1), 1443-1449 (2013)

\bibitem{Tass} M. Khorshid , R.C. Holte, N.R. Sturtevan, A polynomial-time algorithm for non-optimal multi-agent path finding, \textit{The Fourth Annual Symposium on Combinatorial Search}, 76-83 (2011)

\bibitem{PMT} A. Krontiris, R. Luna, K.E. Bekris, From Feasibility Tests to Path Planners for Multi-Agent Pathfinding, \textit{Symposium on Combinatorial Search (SoCS)} (2013)

\bibitem{pafp} P. Kolman, O. Pangrac, On the complexity of paths avoiding forbidden pairs, \textit{Discrete Applied Mathematics 157}, 2871–2876 (2009)

\end{thebibliography}

\end{document}